\documentclass[%
 aip,
 amsmath,amssymb,
 reprint,%
]{revtex4-1}

\usepackage{dcolumn}% Align table columns on decimal point

\usepackage{psfrag}
\usepackage{color}
\usepackage{graphicx,float,amssymb}
\usepackage[bf,loose,normalsize,FIGTOPCAP]{subfigure}
\usepackage{color}
\usepackage{graphicx}
\usepackage{natbib}
\usepackage{bm}
\usepackage{amsmath}

\usepackage[colorlinks = true,
            linkcolor = blue,
            urlcolor  = blue,
            citecolor = blue,
            anchorcolor = blue]{hyperref}

\usepackage{subfigure}
\usepackage{wrapfig}

\begin{document}

%\title{Alignement and Tumbling of Anisotropic Particles\\ in Two-Dimensional Convective Turbulence}
%\title{Preferential Alignment and Tumbling of Anisotropic Particles\\ in Two-Dimensional Convective Turbulence}
\title{Anisotropic particles in two-dimensional convective turbulence}

\author{Enrico Calzavarini}
\affiliation{Univ. Lille, Unit\'e de M\'ecanique de Lille, J. Boussinesq, UML EA 7512, F 59000 Lille, France} 
\email{enrico.calzavarini@polytech-lille.fr}
\author{Linfeng Jiang}
\author{Chao Sun}
\affiliation{Center for Combustion Energy, Key Laboratory for Thermal Science and Power Engineering of Ministry of Education,
Department of Energy and Power Engineering, Tsinghua University, Beijing, China}
\date{\today}

\begin{abstract}
The orientational dynamics of inertialess anisotropic particles transported by two-dimensional convective turbulent flows display a coexistence of regular and chaotic features.  We numerically demonstrate that very elongated particles (rods) align preferentially with the direction of the fluid flow, i.e., horizontally close to the isothermal walls and dominantly vertically in the bulk. This behaviour is due to the the presence of a persistent large scale circulation flow structure, which induces strong shear at wall boundaries and in up/down-welling regions. The near-wall horizontal alignment of rods persists at increasing the Rayleigh number, while the vertical orientation in the bulk is progressively weakened by the corresponding increase of turbulence intensity.
Furthermore, we show that very elongated particles are nearly orthogonal to the orientation of the temperature gradient, an alignment independent of the system dimensionality and which becomes exact only in the limit of infinite Prandtl numbers.  
Tumbling rates are extremely vigorous adjacent to the walls, where particles roughly perform Jeffery orbits. This implies that the root-mean-square near-wall tumbling rates for spheres are much stronger than for rods, up to $\mathcal{O}(10)$ times at $Ra\simeq 10^9$.   
In the turbulent bulk  the situation reverses and rods tumble slightly faster than isotropic particles, in agreement with earlier observations in two-dimensional turbulence. 
\end{abstract}

\maketitle

\section{Introduction}
The rotational dynamics of small anisotropic material particles (\textit{e.g.} rods  or disks)  in turbulent flows has been the focus of a series of recent studies, see  \cite{VothARFM2017} for a review. 
Few state-of-the-art experiments \cite{ParsaPRL2012,ParsaPRL2014,Marcus2014,Byron2015,ni_kramel_ouellette_voth_2015,BounouaPRL2018} as well as several numerical simulations and theoretical studies \cite{chevillard_meneveau_2013,Gustavsson2014,ni_ouellette_voth_2014,CandelierPRL2016,pujara_variano_2017,GustavssonPRL2017} have highlighted their complex behaviour, which is in part inherited from the non-trivial dynamics of the velocity gradient tensor along lagrangian trajectories in developed turbulence. Preferential alignments of particles with intrinsic orientations of the small scale turbulence structures have been observed. For exemple prolate particles preferentially align with the vorticity direction \cite{Gustavsson2014} which tends also to be in line with the second eigenvector of the rate of strain tensor \cite{chevillard_meneveau_2013}. On the opposite oblate particles are mostly orthogonal to such a direction and as a consequence they tumble much faster than rod-like ones \cite{ParsaPRL2012}. 
However, while the phenomenology of orientations is now clear for homogeneous and isotropic turbulent flows (at least for particles of weak inertia), much less explored remains the case of non-homogeneous turbulent flows \cite{VothARFM2017}. Steps in this direction have been made for prolate particles evolving in mixing layers and jets\cite{Lin2003,Lin2012},  in turbulent pipe and channel flows \cite{Lin2005,MarchioliPF2010,Marchioli2013,Zhao2015,Challabotla2015}  and more recently in high-Reynolds number Taylor-Couette flow where the evolution of rigid fibers has been experimentally tracked \cite{Bakhuis2019}.  In the present study we extend the investigation of anisotropic particle dynamics to the paradigmatic case of  turbulent convection in the Rayleigh-B\'enard (RB) system, which displays both an inhomogeneous and anisotropic flow. The present study represents a first step into the exploration of this complex system and for this reason  we limit the investigation to the case of a two-dimensional convective flow advecting anisotropic particles. A similar simplifying approach has been adopted in the past for other types of flows \cite{ParsaPF2011, GuptaPRE2014}. It is however expected that the effect of the dimensionality of the system affects the statistics of rotations of anisotropic particles, as it has been shown in \cite{GuptaPRE2014} for the case of two-dimensional turbulence as compared to three-dimensional developed turbulence.. 
The present study aims at addressing the following open questions:  i) How is the orientation and the rotation of rods affected by the non-homogeneity of the turbulent convective flow? Specifically, what is the effect of coherent flow structures that characterises a thermal-driven flow, in particular the boundary layer (BL), the thermal plumes and the large scale circulation (LSC)? ii) what are the trends at varying the particle shape aspect-ratio and the turbulence intensity (i.e. the Rayleigh number)? iii) finally in which respect the phenomenology of rod dynamics in a 2D system is different from the one in 3D?

The article is organised as follows: In section \ref{sec:method} we present the methodology adopted in this study, in particular we define the model system and concisely describe the set of numerical experiments that have been carried on. Section  \ref{sec:results} will first present the basic phenomenology of the system. It will then guide the reader through the analysis of preferential alignment, tumbling rate and their dependences on the particle anisotropy and on the level of turbulence in the flow. The conclusions, Sec. \ref{sec:conclusions}, summarizes the main finding of this study, its implications and discuss still open topics and perspectives. In the appendix \ref{sec:appendix}  we provide the derivation of the predictions for the tumbling rate of anisotropic particles in two-dimensions that have been checked against the numerical measurements in this work. 

%%%%%%%%%%%%%%%%%.pdf
\begin{figure}[!hb]
\begin{center}
%\subfigure[]{\includegraphics[width=0.7\columnwidth]{visual_new.pdf}}
%\subfigure[]{\includegraphics[width=0.7\columnwidth]{visual_nematic.pdf}}
\subfigure[]{\includegraphics[width=1.0\columnwidth]{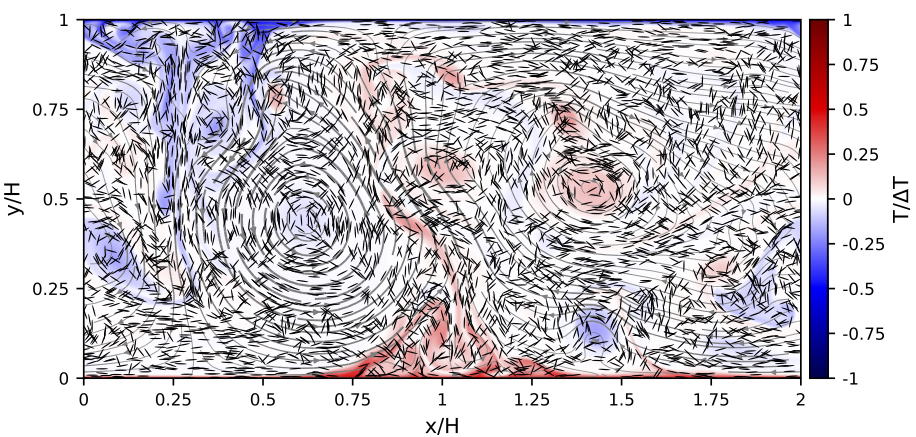}}
\subfigure[]{\includegraphics[width=1.0\columnwidth]{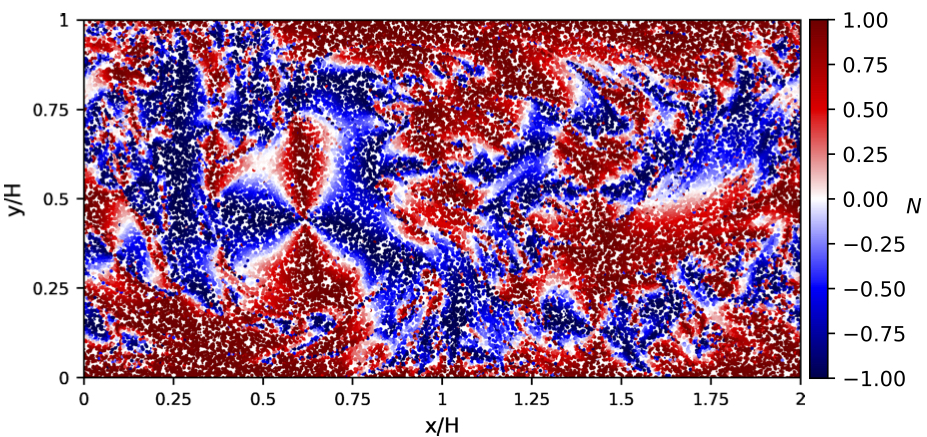}}
\caption{(a) Visualisation of anisotropic particles with aspect ratio $\alpha=100$ in the Rayleigh-B\'enard convective flow at $Ra=10^9$ and $Pr=1$. The size of the particle is arbitrary. The color maps the temperature value, while the grey curves represents the instantaneous flow streamlines. (b) visualisation of the corresponding nematic order parameter $N$. The value 1 (red) indicates horizontal alignment, while -1 (blue) indicates the vertical alignment.}\label{fig:visual}
\end{center}
\end{figure}
%%%%%%%%%%%%%%%%%%

\section{Method \label{sec:method}}
The approach adopted in this study is numerical. We perform a numerical integration of the Boussinesq system of equations,
\begin{eqnarray}
\partial_t \textbf{u} + \textbf{u}\cdot \bm{\partial}   \textbf{u} &=& -   \bm{\partial} p/\rho_0 + \nu\ \partial^2   \textbf{u} + \beta g (T-T_0)  \textbf{\^y} \label{eq:NS}\\
\bm{\partial}  \cdot  \textbf{u} &=& 0 \label{eq:div}\\
\partial_t T + \textbf{u}\cdot \bm{\partial}   T &=&  \kappa\ \partial^2   T, \label{eq:T}
\end{eqnarray}
where $\textbf{u}(\textbf{x},t)$ and $T(\textbf{x},t)$ are respectively the velocity and temperature fields, and the parameters are the kinematic viscosity ($\nu$), the thermal diffusivity ($\kappa$),
the reference density ($\rho_0$) at temperature $T_0$, the thermal expansion coefficient  with respect to the same temperature $(\beta)$ and finally the intensity of gravitational acceleration $(g)$. The domain is rectangular  two-dimensional, with size $H$ in the vertical direction ($y$-axis) and $L=2H$ in the horizontal one ($x$-axis). The boundary conditions on the horizontal planes are no-slip for the velocity, $\textbf{u}=0$,  and isothermal for temperature, $T= T_0 \pm \Delta T/2$, with larger temperature at the bottom wall.  The lateral boundary conditions are periodic for all fields.  The latter choice is made for simplicity in order to have a single direction of statistical non-homogeneity in the flow, i.e., the direction perpendicular to the walls. 
The flow is seeded with point-like anisotropic particles with position, \textbf{r}(t), and orientation, $\textbf{p}(t)$, described by the following set of equations \cite{Jeffery1922}:
\begin{eqnarray}
\dot{\textbf{r}} &=& \textbf{u}(\textbf{r}(t),t)\\
\dot{\textbf{p}} &=& \Omega \textbf{p} + \tfrac{\alpha^2 - 1}{\alpha^2+1} \left( \mathcal{S}\textbf{p} - (\textbf{p}\cdot \mathcal{S}\textbf{p})  \textbf{p} \right) \label{eq:Jeffery3d},
\end{eqnarray}
where $\mathcal{S} = (\bm{\partial}   \textbf{u}+ \bm{\partial}   \textbf{u}^T)/2$ and $ \Omega=(\bm{\partial}   \textbf{u}-\bm{\partial}   \textbf{u}^T)/2$ represent respectively the symmetric and anti-symmetric components of the fluid velocity gradient tensors, $\bm{\partial}   \textbf{u}$, and $\alpha$ is the aspect ratio of the particle assumed to be ellipsoidal and defined as major ($l$) over minor ($d$) axis $\alpha=l/d$. 
In two dimension the orientation equation (\ref{eq:Jeffery3d}) can be conveniently simplified by introducing the orientation angle $\theta$ with respect to the horizontal axis, $\textbf{p}=(p_x,p_y) = (\cos \theta,\sin \theta )$, and taking into account the incompressibility of the flow (see appendix A):
\begin{equation} \label{eq:theta} 
\dot{\theta} =  \frac{1}{2} \omega  -\tfrac{\alpha^2 - 1}{\alpha^2+1} \left[  S_{xx} \sin(2\theta)  - S_{xy} \cos(2\theta) \right].
\end{equation} 
Note that (\ref{eq:Jeffery3d}) is invariant with respect to the transformation $\textbf{p} \to -\textbf{p}$, meaning that it describe fore-and-aft symmetric particles, as a consequence (\ref{eq:theta}) is invariant with respect to the transformation $\theta \to \theta + \pi$.\\ 
We finally note that, when adimensionalized, e.g. by using $H$, $\tau_{\kappa}=H^2/\kappa$, $\Delta T$ as reference scale for length, time and temperature, the above model system has four independent parameters, the Rayleigh number  $Ra=\beta g \Delta T H^3/(\nu \kappa)$, the Prandtl number $Pr=\nu/\kappa$, the geometrical aspect ratio of the domain $\Gamma=L/H$ and of the particle $\alpha$. However, in the forthcoming analysis it will be convenient also to consider as a reference time-scale, the dissipative time scale of the flow, $\tau_{\eta} = \sqrt{\nu/\bar{\epsilon}}$ with $\bar{\epsilon}$ the global mean energy dissipation rate. In the RB system such a time scale can be also expressed as $\tau_{\eta} = \tau_{\kappa}/\sqrt{Ra(Nu-1)}$ where $Nu$ is the mean Nusselt number in the system.

In this study we explore the particle aspect ratio dependency $\alpha$, and the $Ra$ number that  parametrizes the strength of the thermal convection in the flow. 
We evolve $N_p=O(10^5-10^6)$ particles divided into 20 aspect ratio types, logarithmically spaced in the interval $\alpha \in [1,100]$. 
The Rayleigh number spans the range $Ra \in [2.44 \times 10^5, 8 \times 10^9]$.
The simulations are performed through a well tested computational fluid dynamics code, already adopted in a series of previous studies \cite{Calzavarini2019}.  Table \ref{tab:1} reports the relevant control parameters in the numerical simulations. 

\begin{table}[!htb]
\begin{center}
\begin{tabular}{c | c | c | r }
Ra & $N_x \times N_y$ & $\tau / \tau_H$ & $N_p \qquad $\\
\hline
$2.44 \times 10^5$ & $128 \times 64$ & 280 & $1.25 \times 10^5$ \\
$1.95 \times 10^6$ & $256 \times 128$ & 242 & $1.25 \times 10^5$ \\
$1.56 \times 10^7$ & $512 \times 256$ & 159 & $1.25 \times 10^5$\\
$1.25 \times 10^8$ & $1024 \times 512$ & 188 & $2.5 \times 10^5$\\
$1.00 \times 10^9$ & $2048 \times 1024$ & 135 & $10^6$\\
$8.00 \times 10^9$ & $4096 \times 2048$ &  22 & $4 \times 10^6$\\
\end{tabular}
\end{center}
\caption{Main parameters of the numerical simulations: the Rayleigh number $Ra$; the horizontal ($N_x$) and vertical ($N_y$) size of the grid; the total duration of the simulation $\tau$ in integral turnover time units $\tau_H = H/u_{rms}$; the total number of particles ($N_p$) evolved in each simulation.}\label{tab:1}
\end{table}

\section{Results \label{sec:results}}
We begin with a detailed analysis of the particles alignment and tumbling rate as a function of their aspect-ratios in prescribed flow conditions at $Ra=10^9$ and $Pr=1$. The dependence of these phenomena on the strength of the thermal forcing, parametrised by the $Ra$ number, will be addressed in a separate section.  

\subsection{Preferential alignment}
Figure \ref{fig:visual}(a)  displays a visualisation of an instantaneous configuration of highly anisotropic particles, a set of $5\times10^4$ particles with $\alpha = 100$, together with the fluid flow field streamlines and a heat-map of the temperature field. One can appreciate the fact that the particle orientation is visually correlated to the flow structures. Close to the walls particles appear preferentially horizontal, while along and inside upwelling and downwelling thermal plumes they looks predominantly vertical. Furthermore, they seem to be influenced by the presence of a LSC flow structure, this is evident from the clear tendency to align along streamlines, and to a minor extent by the presence of secondary gyres in the system.

In order to better appreciate the trend displayed by the particles orientation as a function of their local position one can use the nematic order parameter \cite{GuptaPRE2014},
\begin{equation}
N \equiv  2  \ (\textbf{p}\cdot \textbf{\^x})^2  -1 = 2 (\cos \theta)^2   - 1, 
\end{equation}
which takes the value 1 in case of a perfect horizontal alignment along the x-axis, and -1 if the particles are vertically aligned. The visualisation, provided in figure \ref{fig:visual}(b), shows the local value of $N$ for the same instant of time presented in panel (a). It is now more evident the phenomenon of i) preferential alignment at the wall, ii) the vertical orientation in plume dominated regions and iii) the ordering effect along streamlines produced by energetic vortices.  It is worth noting that, due to the random initial conditions that we adopted for the particle orientations, the quantity $N$ can not be approximated to a smooth field, not even in the long time  \cite{Zhao2019}. This is the reason why in figure \ref{fig:visual}(b) we still observe dots of a different colour inside large domains of particles mostly aligned along the same direction.

\begin{figure}[!ht]
\begin{center}
\includegraphics[width=1.0\columnwidth]{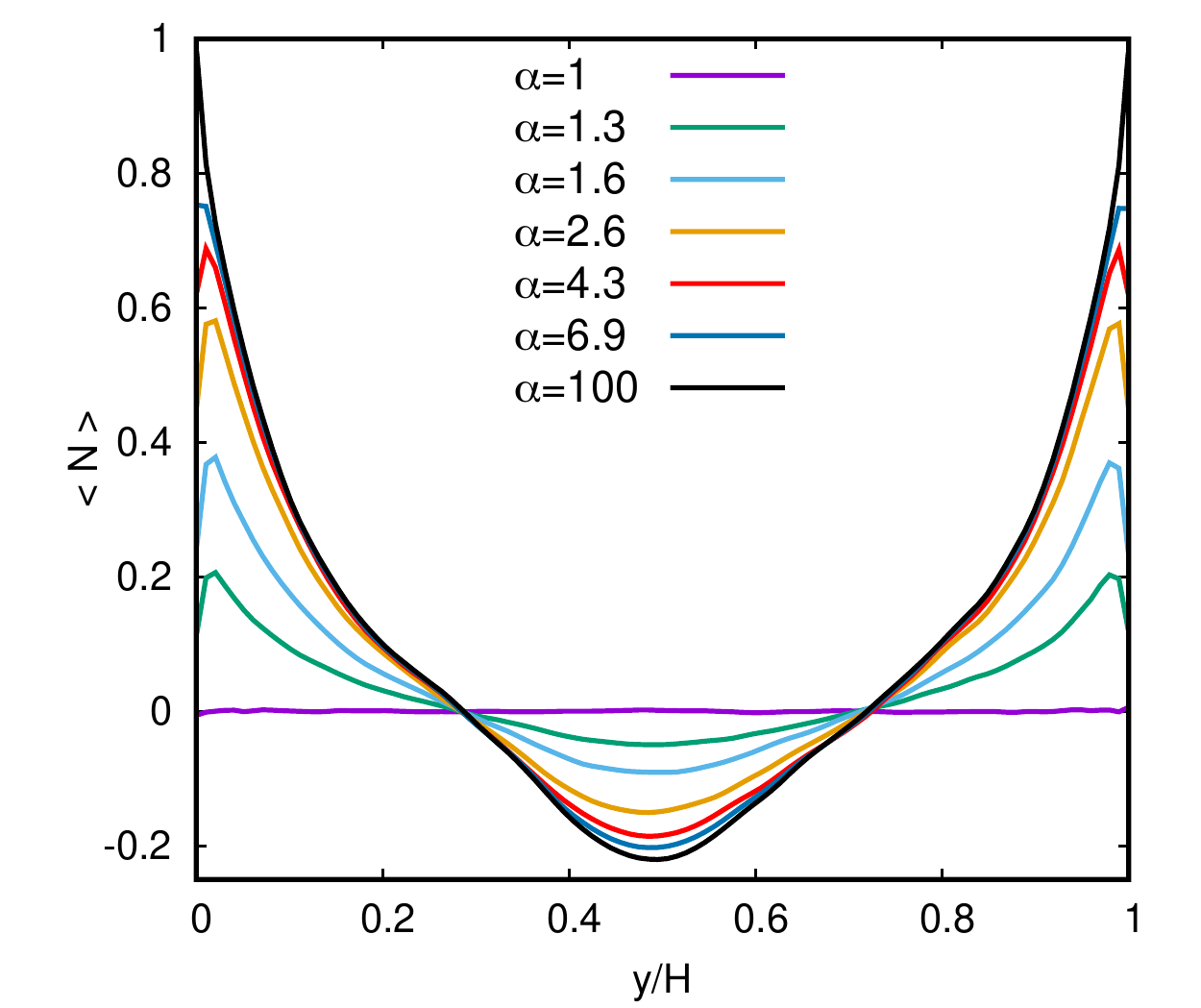}
\caption{Local nematic order parameter as a function the distance from a horizontal wall in the system, for different particles aspect ratios at $Ra=10^9$, $Pr=1$. We compute the average $\langle N \rangle (y)$, where $\langle \ldots \rangle$ is taken over time and over the particles with given $y \pm \delta y$ coordinates, here with $ \delta y = H/2048$.  It is shown that the more anisotropic is the particle, the more it displays a non random orientation. For high values of $\alpha$ the alignment is nearly perfectly parallel to the wall in flow regions close to the wall, while in the bulk a clear tendency to be perpendicular to the walls is observed.}\label{fig:nematic}
\end{center}
\end{figure}

In order to quantitatively appreciate the mean trend displayed by the orientation as a function of the position in the system, specifically the distance from the walls, and at the same time as a function of the aspect ratio of the particles, we compute the average $\langle N \rangle (y)$, where $\langle \ldots \rangle$ is taken over time and over the particles with given $y \pm \delta y$ coordinates. The interpretation of the mean value of the nematic order parameter over a given region of space is slightly different, than its instantaneous value, while the meaning of the limiting cases $\pm 1$ remains the same, the zero value is likely to indicate a statistically isotropic distribution (given the unsteady nature of the flow   the case in which all particles at a given height are oriented on $\pm 45 \deg$ angle is unlikely). Figure \ref{fig:nematic} shows how the mean orientational ordering varies as a function of the distance from the top and bottom walls, that is to say in the direction of inhomogeneity in the flow. One can note the symmetry of the curves with respect to the mid plane which attests the excellent convergence of the simulations. 
For the isotropic particles, $\alpha = 1$, as expected there is no preferential orientation and  $\langle N \rangle =0$ at any level. However, as soon as the shape anisotropy comes into play particles tend to align preferentially horizontally next to the walls and weakly vertically in the bulk of the system. It appears that, at the considered $Ra$, for all particle anisotropic classes the statistically random orientation region occurs at roughly one third of the box height. In the case of highly anisotropic particles ($\alpha = 100$) the orientation is nearly perfectly horizontal at the system boundaries. We stress that this remarkable effect can not be related to a direct interaction of the rods with the walls (wall-rods collisions are not implemented in our model system) but it is rather a dynamical effect mediated by the properties of the fluid gradient at the particle position in that region of the domain. We will come back later on this important feature.

%%%%%%%%%%%%%%%%%
\begin{figure}[!ht]
\begin{center}
\includegraphics[width=1.0\columnwidth]{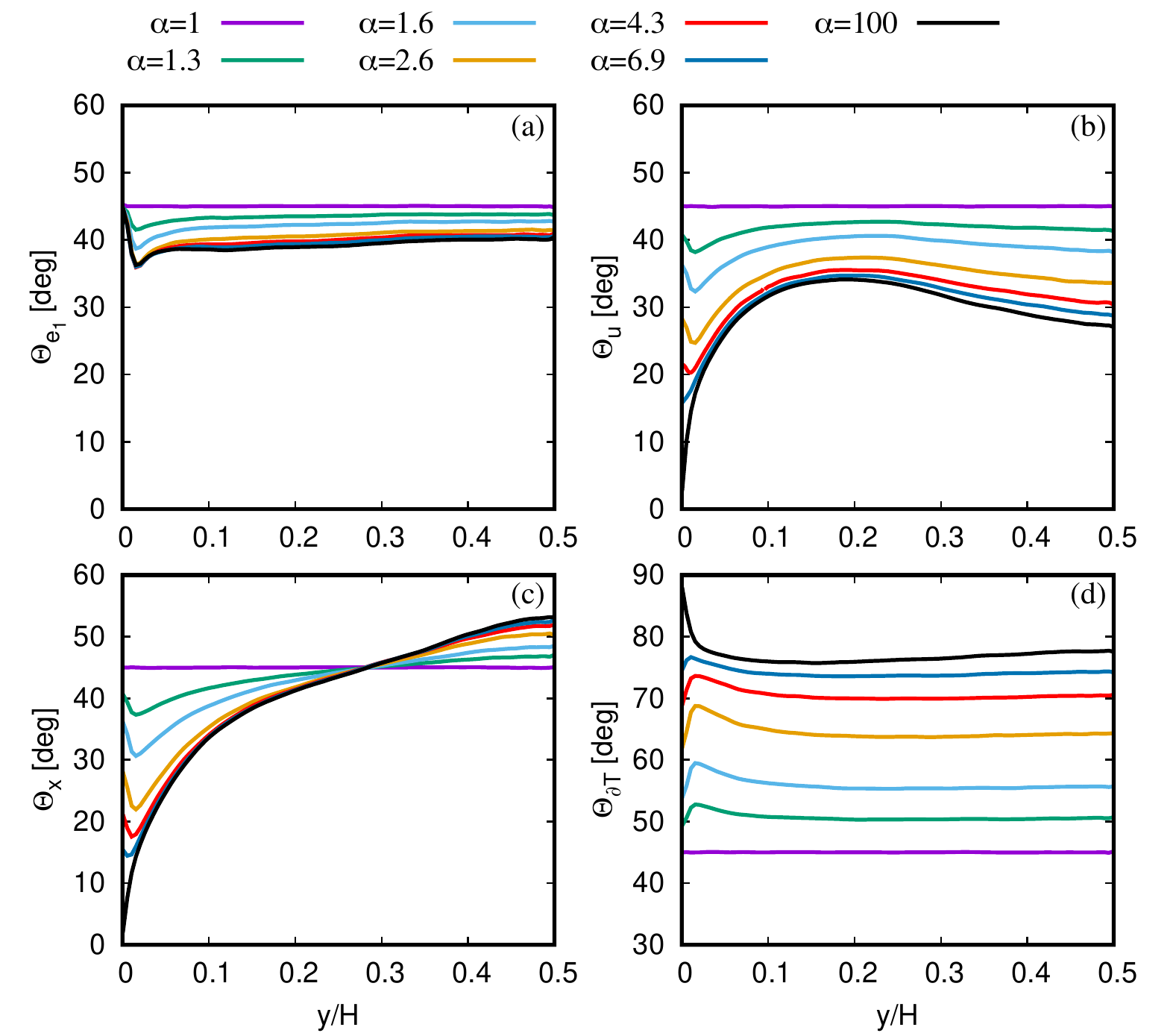}
\caption{Mean orientation angle with respect to the first eigenvector of the rate-of-strain tensor $\textbf{e}_1$ (a); the fluid velocity vector $\textbf{u}$ (b); the $x$ axis (c); the temperature gradient $\bm{\partial}T$ (d), for various particle aspect ratios ranging from spheres $\alpha=1$ to rods $\alpha=100$. $Ra=10^9$, $Pr=1$.}\label{fig:orientation}
\end{center}
\end{figure}

%%%%%%%%%%%%%%%%%%
So far we have observed that the particles preferentially align along the cartesian axis of the system. However, since the particle do not interact directly with the wall boundaries, this must be a consequence of the structure of the flow field in the system.  In order to better understand this aspect we  measure the average orientation angle of the particle with respect to a given vector $\textbf{a}$, this is done by taking 
$$\Theta_a = \langle \arccos{ \left| \textbf{p} \cdot \frac{\textbf{a}}{|| \textbf{a}||} \right|} \rangle,$$
where is to be noted that $\Theta_a \in [0,\pi/2]$ due to the fore-aft symmetry of the particles.
We consider the cases in which the $\textbf{a}$ vector is again the horizontal direction (x-axis) but also the fluid velocity $\textbf{u}$, the eigenvector $\textbf{e}_1$ corresponding to the largest eigenvalue of the strain rate tensor $\mathcal{S}$, and the temperature gradient $\mathbf{\partial} T$.
The results reported in figure \ref{fig:orientation} illustrates the behaviour of the mean angle at increasing the distance from the wall. The overall strongest alignment is found for highly anisotropic particles with the direction of the flow, $\textbf{u}$ ( fig. \ref{fig:orientation}(a)). We note that such an alignment is very strong near to the system boundaries, where the velocity is mostly parallel to the x axis (see fig. \ref{fig:orientation}(b) ), but the alignment remains noticeable also in the bulk, where the velocity has a dominant vertical component. 

On the contrary, the alignment of the anisotropic particles with $\textbf{e}_1$ is weak, fig. \ref{fig:orientation}(c), a feature that was already observed in the case of homogeneous 2D turbulent flow \cite{GuptaPRE2014}. This can be also understood by reformulating eq. (\ref{eq:theta}) in terms of the angle $\theta_1$ formed by $\textbf{e}_1$ with the x-axis.
This gives (see \ref{sec:appendix}):
\begin{equation}
\dot{\theta} =  \frac{1}{2} \omega  -\frac{\alpha^2 - 1}{\alpha^2+1}\sqrt{ S_{xx}^2 + S_{xy}^2}\ \sin(2(\theta-\theta_1)).
\end{equation}  \label{eq:theta1} 
If vorticity was absent the above equations would have a fixed point $\theta = \theta_1 + n \pi/2 $ with $n=0,1$, independently of the aspect ratio. 
This means that both the alignment with $\textbf{e}_1$ or with the orthogonal eigenvector $\textbf{e}_2$ are equally favoured. 
However, the presence of vorticity, which is moreover local and time dependent, inevitably perturbs and removes such equilibrium positions. 

Another salient aspect is the nearly orthogonal alignment of rodlike particles with the local temperature gradient  (fig. \ref{fig:orientation}(d) ). This feature is related to the fact that the equation for the temperature gradient orientation $\hat{\bm{\partial}T} = \bm{\partial}T / || \bm{\partial}T ||$ shares  similarities with the one of anisotropic particles. One has
\begin{equation}
\dot{\hat{\bm{\partial}T}}  =   \Omega \hat{\bm{\partial}T}  - \mathcal{S} \hat{\bm{\partial}T}   + (\hat{\bm{\partial}T})^T \mathcal{S}\hat{\bm{\partial}T}  \hat{\bm{\partial}T} + \mathcal{O}(\kappa), 
\end{equation}
where $\mathcal{O}(\kappa)$ denotes the diffusive terms that are linear in $\kappa$.
It is possible to show that, when the diffusive terms are neglected, the unit vector $\hat{\bm{\partial}T}$ follows the same evolution of a vector orthogonal to $\textbf{p}$ for $\alpha \to \infty$ (see \ref{sec:appendix}).  
This means that, in a statistical sense, and when the effect of thermal diffusion is negligible (limit of large Prandtl number) the orientation of $\bm{\partial}T$ shall be orthogonal to the one of thin rods.  To our knowledge this phenomenon has never been reported or tested before. We also note that such an analogy is independent of the dimensionality and therefore it must hold also in 3D (in the 3D case the orientation of very oblate particles, disks, will preferentially align along the thermal gradient direction). The origin of this alignment is analogous to to the one that exists between the equation for the vorticity director and the Jeffery equation for a thin rod ($\alpha\to \infty$) \cite{PumirWilkinson2011}.

%%%%%%%%%%%%%%%%%
%%%%%%%%%%%%%%%%%%

\subsection{Tumbling rate}
\begin{figure}[!hb]
\begin{center}
%\subfigure[]{\includegraphics[width=0.7\columnwidth]{visual_tumbling_sphere.pdf}}
%\subfigure[]{\includegraphics[width=0.7\columnwidth]{visual_tumbling_rod.pdf}}
\subfigure[]{\includegraphics[width=1.0\columnwidth]{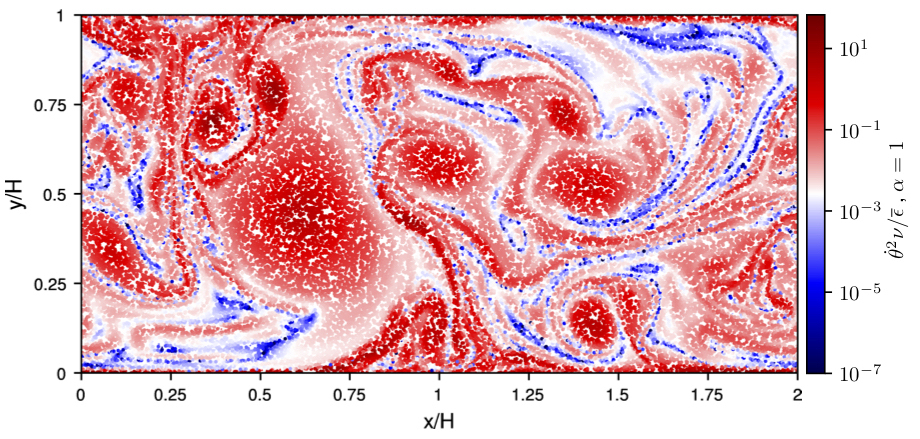}}
\subfigure[]{\includegraphics[width=1.0\columnwidth]{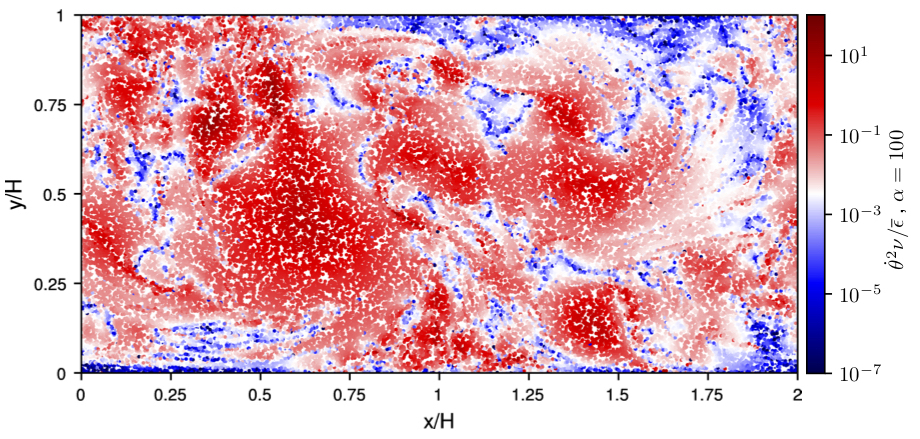}}
\caption{Visualisation of the instantaneous local value of quadratic tumbling rate for isotropic $\alpha=1$ (a) and highly elongated particles $\alpha=100$ (b). The flow conditions are $Ra=10^9$ and $Pr=1$, the instant of time (and correspondingly the flow field) is the same as in Fig. \ref{fig:visual}.
%Snapshot of anisotropic particles with aspect ratio $\alpha=1$ (a)  and $\alpha=100$ (b) in the Rayleigh-B\'enard convective flow at $Ra=10^9$ and $Pr=1$. The color reports the temperature values, while the grey curves represents the instantaneous flow streamlines.
}\label{fig:visual-tumbling}
\end{center}
\end{figure}

\begin{figure}[!ht]
\begin{center}
\subfigure[]{\includegraphics[width=1.0\columnwidth]{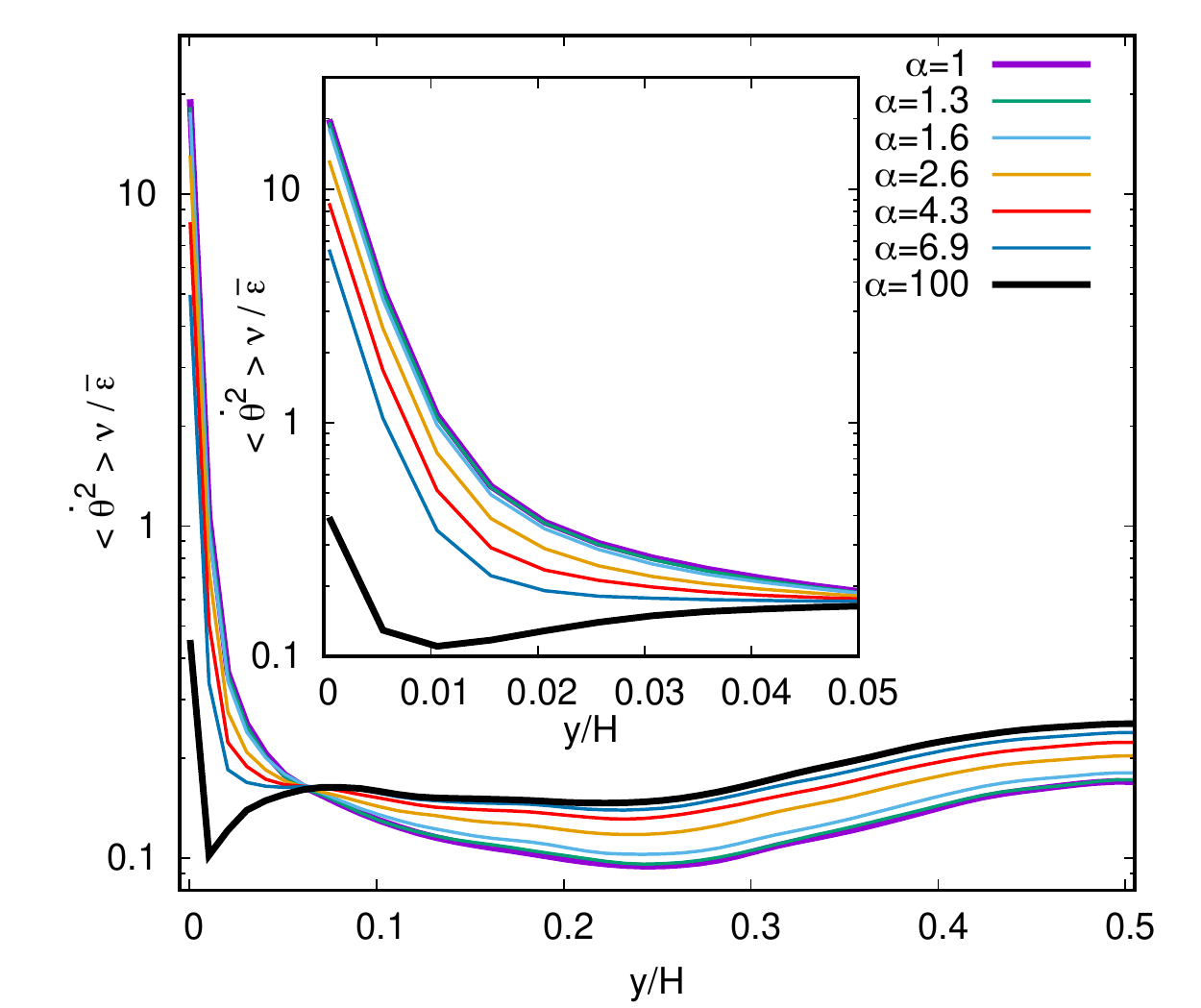}}
\subfigure[]{\includegraphics[width=1.0\columnwidth]{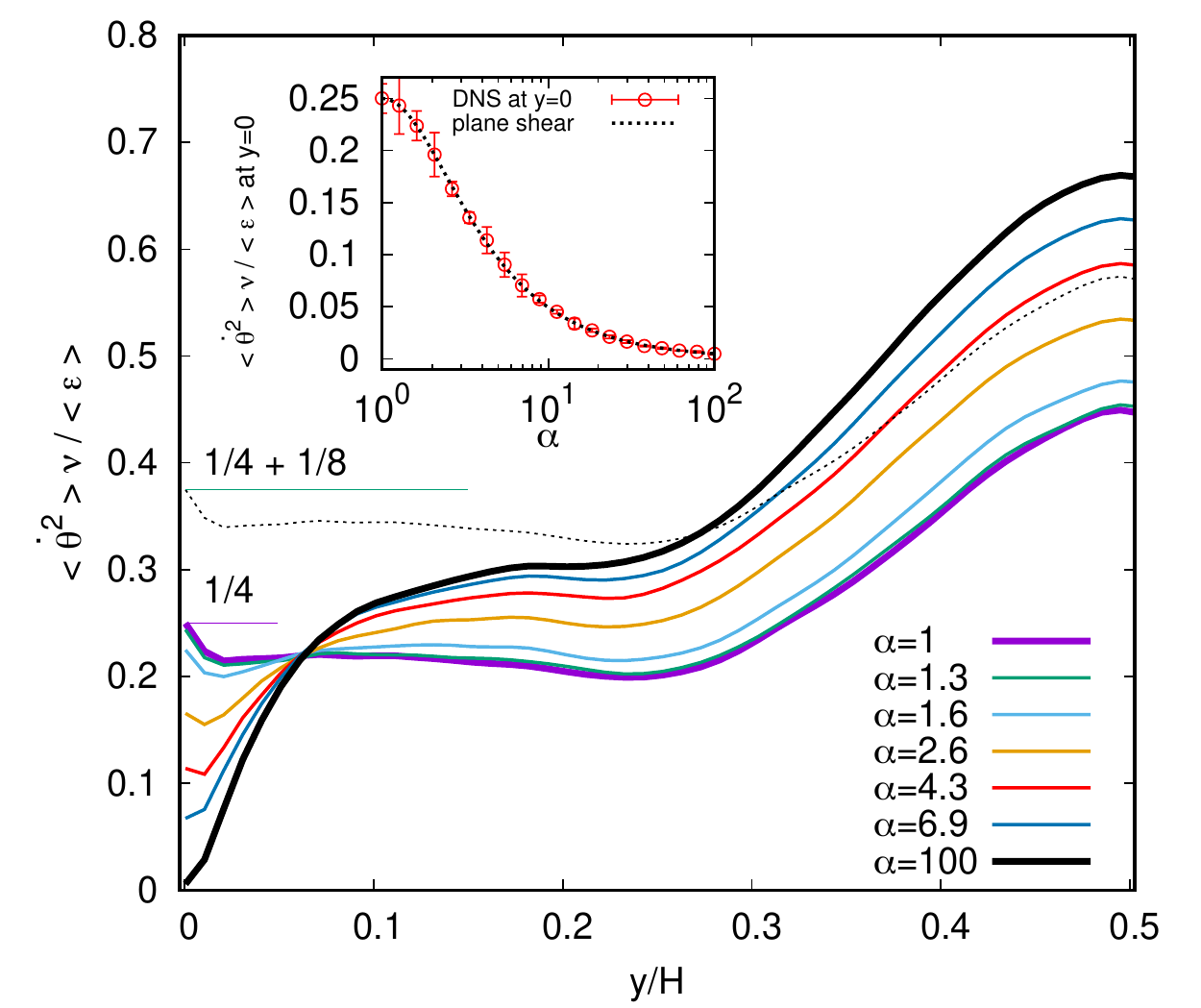}}
\caption{(a) Mean quadratic tumbling rate, $\langle \dot{\theta}^2 \rangle $ as a function of the distance from the wall $y\in [0,H/2]$ for different particle aspect ratios. The tumbling rate is normalized by means of the global energy dissipation rate $\overline{\epsilon}$. The inset reports a zoomed-in vision of the the wall region.
(b) Same as before but with a normalization based on the local dissipative energy dissipation rate $\langle \epsilon \rangle$. The dotted line reports the no-correlation prediction (\ref{hyp:random}) for $\alpha=100$, the continuous horizontal lines gives the values of the isotropic flow  prediction (\ref{hyp:iso}) for $\alpha=1$ (minimum value) and $\alpha=100$(maximum value). The inset reports the values (datapoints) of the normalized quadratic tumbling rate at the wall ($y = 0$) and a comparison with the prediction (\ref{jeffery-tumbling}), which describe the tumbling in a plane shear flow, also named Jeffery tumbling.}
\label{fig:tumbling-rate}
\end{center}
\end{figure}

The observations made in the previous section can be further supported by means of the study of the rotation rate of the particles. Because this rotation is around an axis orthogonal the particle symmetry direction, it is common to name it tumbling.  We study here the quadratic tumbling rate intensity, which can be expressed in terms of the quantity $ \dot{\textbf{p}} \cdot  \dot{\textbf{p}} = \dot{\theta}^2$. 
First we visualise the instantaneous value of  such a quantity both for isotropic $\alpha=1$ and highly elongated particles $\alpha=100$, see  Fig. \ref{fig:visual-tumbling}.
Note that the quadratic tumbling rate of isotropic particles is by definition proportional to the local fluid vorticity, via $\dot{\theta}^2 = \omega^2 / 4$. As a result we see that $\alpha=1$ particles tumble vigorously near to walls, where the vorticity is generated and close to vortex cores.  The elongated particles clearly tumble much less at the walls, but show a similar tumbling rate distribution in the bulk, maximal inside vortices although smeared down as compared to the case of spheres.\\  
%%%%%%%%%%%%%%%%%
%%%%%%%%%%%%%%%%%%
Such qualitative differences are again better understood by looking at their mean behaviours.  Figure \ref{fig:tumbling-rate} shows the mean quadratic tumbling rate (we use the average $\langle \ldots \rangle$ with the same meaning as before) normalized by the squared global dissipative time-scale, i.e. $\overline{\epsilon}/\nu$ where $\epsilon = 2 \nu \mathcal{S}:\mathcal{S}$. Although this normalization is not the most suitable for such a type of flow, which is strongly inhomogeneous, it has the advantage to allow for a direct comparison of the intensity of tumbling among different vertical positions.
Indeed, here we clearly observe that the tumbling rate is exceptionally high in the boundary layer. Furthermore, it is much higher for spheres as compared to rods. This hierarchy is reverted in the bulk of the flow, where rods tumble slightly faster than spheres, Fig.\ref{fig:tumbling-rate}(a).  
What happens in the bulk of the flow? As we mentioned in the introduction, in three dimensional turbulence anisotropic particles develops correlations with the flow gradient and as a result the mean tumbling rate has a peculiar behaviour as a function of the aspect ratio of the particles. In particular, prolate particles ($\alpha > 1$) shows a rapid decrease of mean tumbling rate, $(\langle \dot{\textbf{p}} \cdot  \dot{\textbf{p}} \rangle)^{1/2}$ for increasing $\alpha$ and a saturation occurring at around $\alpha \simeq  5$ to a value which about 80\% less then the root-mean-square tumbling rate for spherical particles \cite{ParsaPRL2012}. In two dimensional turbulence such an effect has been reported to revert \cite{GuptaPRE2014}. A smooth increase of tumbling with $\alpha$  has been observed in 2D, although with a dependence on the type of forcing applied to sustain the turbulent flow. 

In order to better understand the phenomenology of rotation it is particularly useful to adimensionalize  the quadratic tumbling rate at a given distance from the wall by the time scale based on the local energy dissipation rate $\langle \epsilon \rangle/\nu$, which is shown in Fig. \ref{fig:tumbling-rate}(b). It appears that with this rescaling the rotation rate at the wall for spheres is close to the value 1/4 while for anisotropic particles it tends to vanish. This feature is explained by taking into account that close to the walls the shear term $\dot{\gamma} = \partial_y u_x$  is the dominant one. If we assume it to be time and space (along x direction) independent and we plug it into the Jeffery equation, one gets the prediction for the tumbling rate in the case in which particles are performing the so called Jeffery orbits (see \ref{sec:appendix} for a derivation):
\begin{equation}\label{jeffery-tumbling}
  \frac{\langle \dot{\theta}^2  \rangle}{\langle\epsilon\rangle/\nu}  =  \frac{\alpha}{2(\alpha^2+1)}
\end{equation}
We observe that in the isotropic limit, $\alpha = 1$, one gets $1/4$ while in the very elongated case the rotation rate vanishes. This simplified model prediction is in excellent agreement with the simulations  (see the inset of \ref{fig:tumbling-rate}(b)). It is indeed known that Jeffrey orbits of prolate particles are characterized by a non-uniform tumbling velocity that reaches its minimum when the particle orientation is along the streamlines (and is maximal in the the shear direction) \cite{Jeffery1922}. This phenomenon is responsible for producing the observed alignment of particles in near-wall regions. 
%%%%%%%%%%%%%%%%%
%%%%%%%%%%%%%%%%%%

\begin{figure}[!ht]
\begin{center}
\includegraphics[width=1.0\columnwidth]{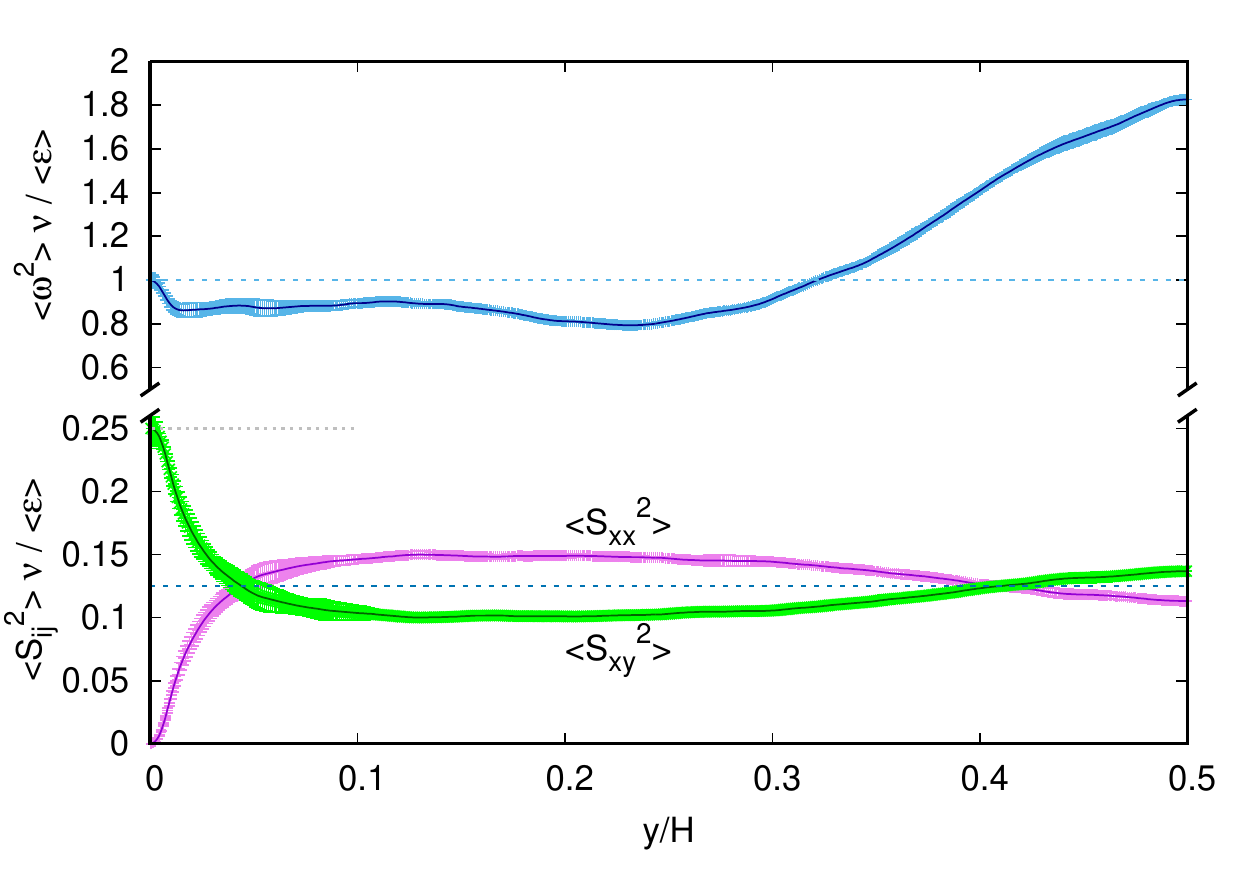}
\caption{Check of local small-scale flow isotropy: The continuous lines represent $\langle \omega^2 \rangle$,$\langle S_{xx}^2 \rangle$ and $\langle S_{xy}^2 \rangle$ in $\langle \epsilon \rangle / \nu$ units (i.e. local dissipative units) as a function of the distance from the wall $y \in \left[ 0,H \right]$. The colour shadow around the lines indicates the standard deviation error bars.
The dashed lines provides the values expected in the isotropic case, $\langle \omega^2 \rangle \nu / \langle \epsilon \rangle = 1$ and $\langle S_{xx}^2 \rangle \nu / \langle \epsilon \rangle =  \langle S_{xy}^2 \rangle \nu / \langle \epsilon \rangle = 1/8$.   The dotted line reports the value expected for plane shear flow, when the only non-null velocity gradient component is $\partial_y u_x$. $Ra=10^9$, $Pr=1$.}
\label{fig:isotropy}
\end{center}
\end{figure}

A quantitative prediction might be attempted also for the bulk of the system along the following lines. One can assume that in turbulent regime the i) fluid velocity gradient components are statistically independent and ii) that they are uncorrelated with the particle orientation angle. This leads  to:
\begin{equation}\label{hyp:random} 
\langle \dot{\theta}^2  \rangle =  \frac{1}{4}\langle \omega^2 \rangle + \frac{1}{2}\left(\frac{\alpha^2 - 1}{\alpha^2+1}\right)^2 \left[  \langle S_{xx}^2 \rangle   +  \langle S_{xy}^2 \rangle \right]
\end{equation}
The further assumption iii) of statistical isotropy of the flow (see appendix \ref{sec:appendix}) leads to:
\begin{equation}\label{hyp:iso} 
\frac{\langle \dot{\theta}^2  \rangle}{\langle \epsilon \rangle /\nu} =  \frac{1}{4} + \frac{1}{8} \left(\frac{\alpha^2 - 1}{\alpha^2+1}\right)^2.
\end{equation}
Note that the above expressions correctly predicts an increase of the quadratic tumbling rate with $\alpha$. However, the predicted tumbling rate appear to be quite off from what is observed in the bulk of the system, see Fig. \ref{fig:tumbling-rate} (b). The reason of this offset can be principally ascribed by the assumption of statistical isotropy  of the flow. Indeed, a direct test of isotropy, reported in 
Fig. \ref{fig:isotropy} shows the net dominance of the vorticity in the bulk of the system.
A direct comparison of eq. (\ref{hyp:random}) with the measurements capture the correct trend for the tumbling in the bulk. This is reported Fig. \ref{fig:tumbling-rate} (b) where the dotted line corresponds to eq. (\ref{hyp:random}) for $\alpha=100$.
Note that the \textit{no-correlation} prediction obviously fails in the BL and near wall regions, where the already discussed Jeffery tumbling occurs. 
%%%%%%%%%%%%%%
%%%%%%%%%%%%% 

\subsection{Rayleigh number dependence}
How general is the description we have provided so far? In this section we examine the dependence of our findings with respect to the level of turbulence in the system. In order to do so we compare the averaged nematic orientations and quadratic tumbling rates of isotropic $\alpha=1$ and highly anisotropic particle $\alpha=100$ at varying the $Ra$ number of the flow. This is obtained by means of numerical simulations in which all the simulation parameters are kept the same except the size of the bounding box. We explore the range $Ra \in \left[2.4\times 10^5,8 \times10^9 \right] $. 

%%%%%%%%%%%%%%%%%
\begin{figure}[!ht]
\begin{center}
\includegraphics[width=1.0\columnwidth]{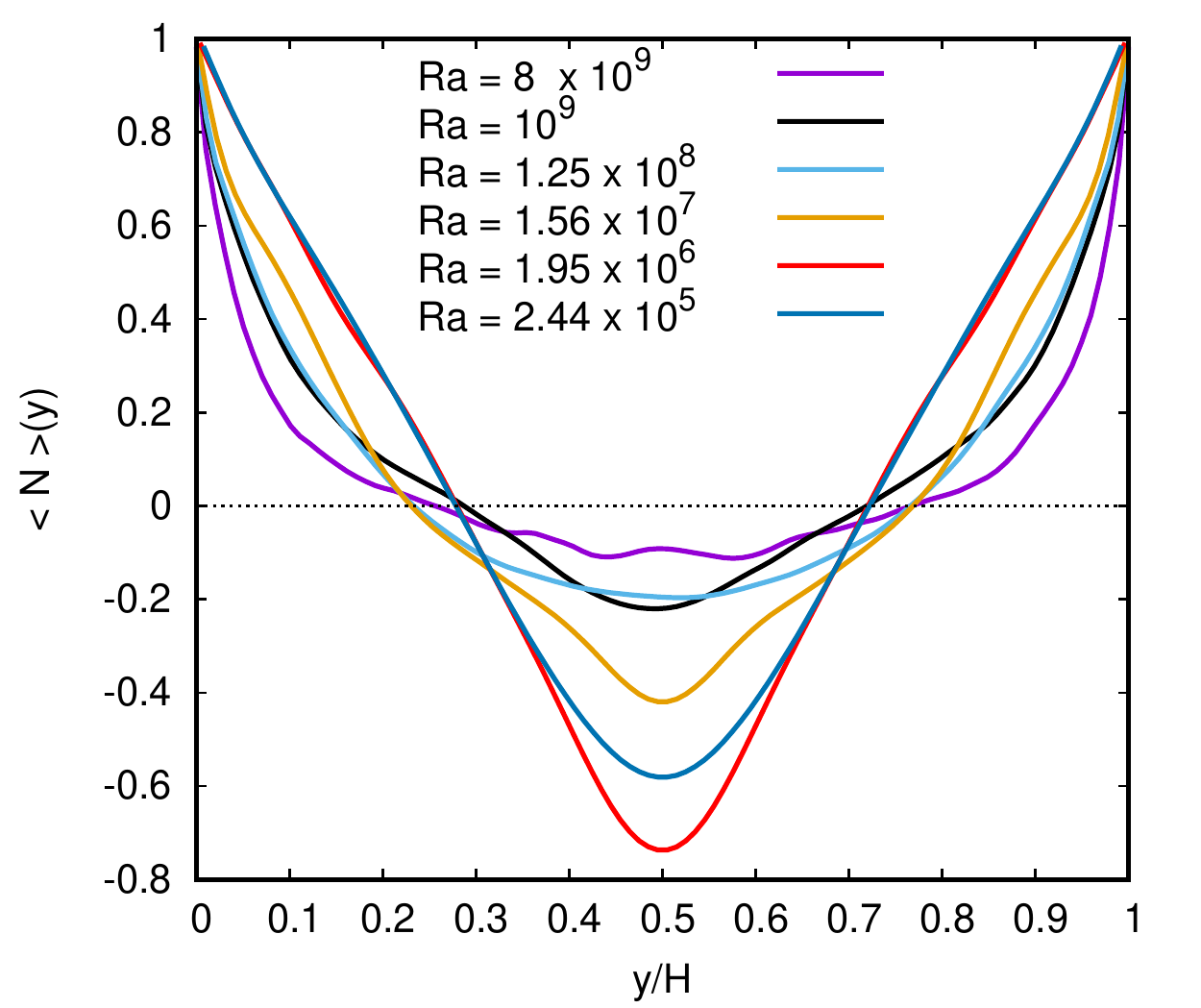}
\caption{Local nematic order parameter as a function the distance from a horizontal wall in the system, for particles of aspect ratios $\alpha =100$ and for Rayleigh numbers $Ra \in \left[2.4\times 10^5,8 \times10^9 \right] $.}\label{fig:nematic_Ra}
\end{center}
\end{figure}
%%%%%%%%%%%%%%%%%%

The results on nematic ordering  for the most anisotropic particles $\alpha = 100$ are reported in fig \ref{fig:nematic_Ra}. One can appreciate that the wall alignment at the system boundaries is a persistent feature at any $Ra$ number. However, larger values of $Ra$ produce a thinning of such regions, which probably reflects the thinning of kinetic BL. The bulk region of the system tends to loose any trace of preferentially vertical alignment and get closer to a value $\langle N \rangle \sim 0$ indicating an isotropization of the orientation. Indeed the mechanism leading to the vertical alignement in the bulk for highly anisotropic particles is the same occurring for the horizontal alignement at the walls, i.e. plane-shear dominated tumbling occuring at the edge of large scale circulations cells where up- or down-welling occurs. The regularity of LSC is weakens with the increase of $Ra$ and so the observed vertical alignement in the bulk. 

\begin{figure}[!ht]
\begin{center}
\subfigure[]{\includegraphics[width=0.87\columnwidth]{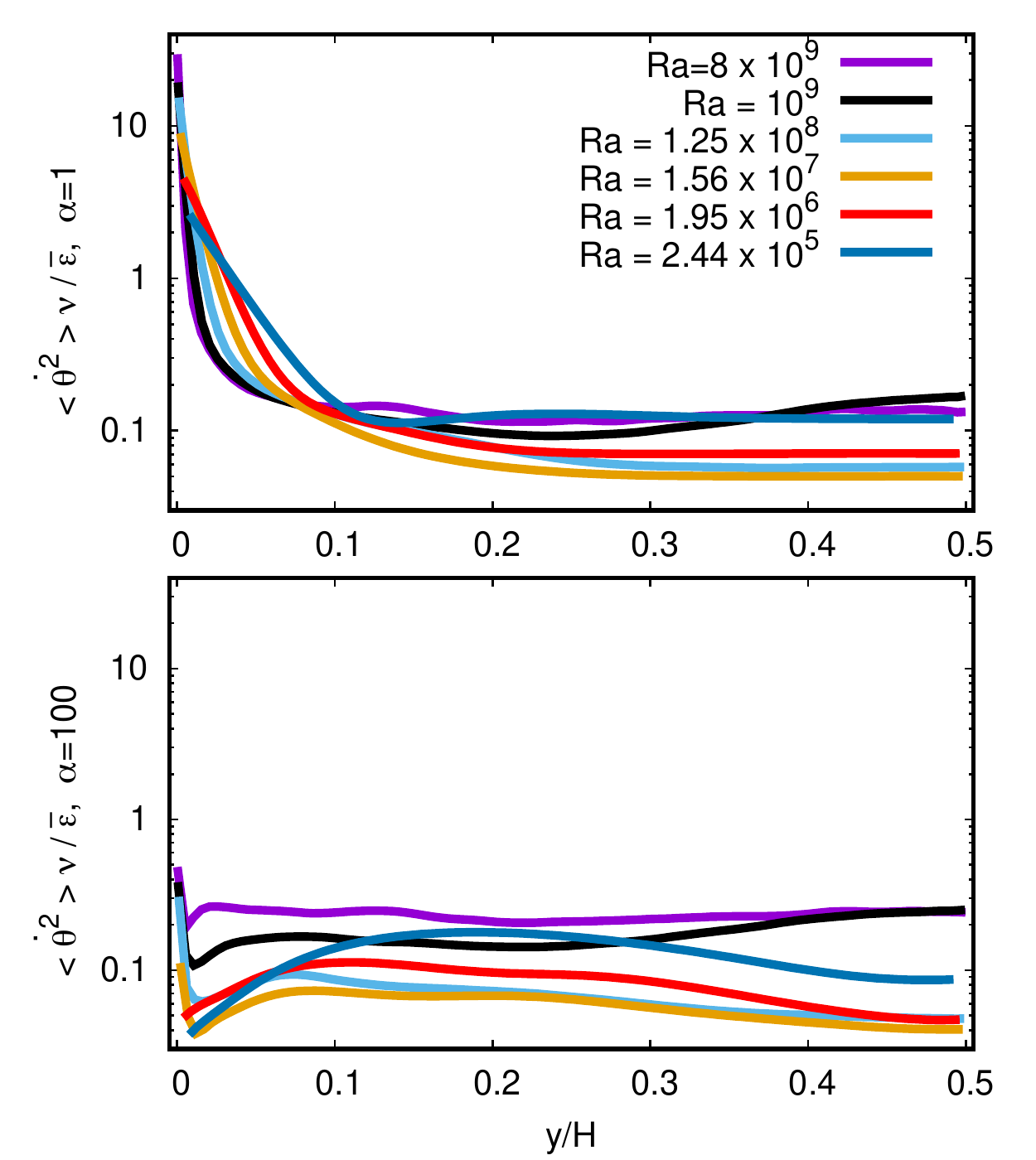}}
\subfigure[]{\includegraphics[width=0.87\columnwidth]{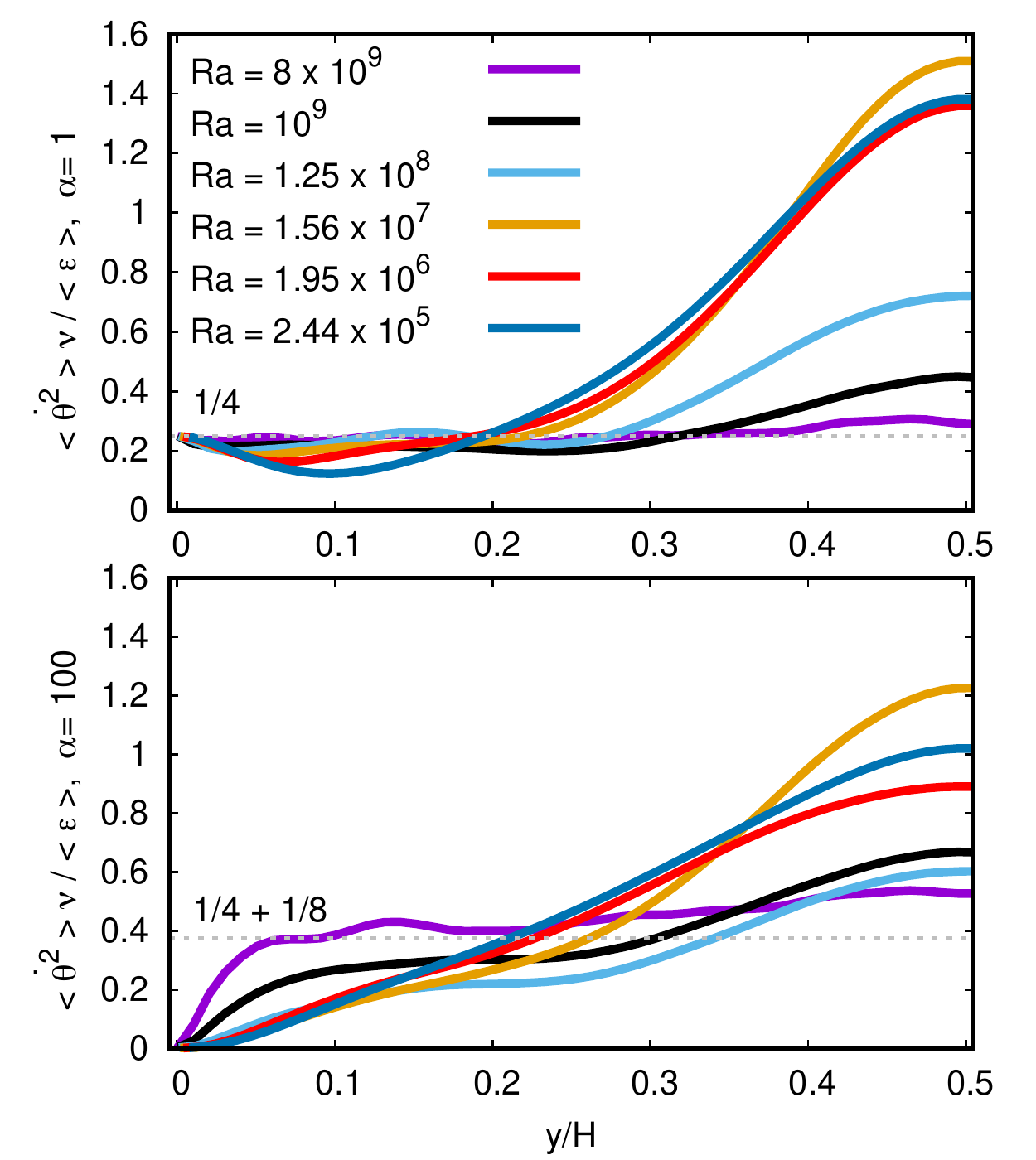}}
\caption{(a) Mean quadratic tumbling rate, $\langle \dot{\theta}^2 \rangle $ as a function of the distance from the wall $y\in [0,H/2]$ at different $Ra$ numbers for $\alpha=1$(top) and $\alpha=100$(bottom) . The tumbling rate is normalized by means of the global energy dissipation rate $\overline{\epsilon}$. 
(b) Same as above but with a normalization based on the local dissipative energy dissipation rate $\langle \epsilon \rangle$. 
The dashed horizontal lines gives the values of the isotropic flow  prediction (\ref{hyp:iso}), respectively 1/4 and 3/8 for $\alpha=1$ and $\alpha=100$.
}\label{fig:tumbling_Ra}
\end{center}
\end{figure}

Figure \ref{fig:tumbling_Ra}(a) reports the tumbling rate for spheres and elongated particles with the global dissipative time normalization at varying $Ra$ numbers. We observe an overall enhancement of tumbling at increasing $Ra$, both in the near-wall and bulk regions. The measurements also confirm the stronger tumbling for spheres than rods for close to the walls. 

Figure \ref{fig:tumbling_Ra}(b) which uses the local energy dissipation rate normalization highlights the attainment of isotropy in the bulk of the system at increasing Rayleigh.
The predictions (\ref{hyp:iso}) based on the decorrelation with the gradient and isotropization are nearly satisfied for the highest $Ra$ simulated.  The system isotropization at the highest $Ra$ is independently confirmed by a direct check of isotropy (see Additional Material). This measurement confirms that in the asymptotic $Ra$ limit rods will tumble slightly more than spheres in the bulk of the system, just the opposite trend as compared to rods in 3D turbulent flows.

 %%%%%%%%%%%%%%%%%
%%%%%%%%%%%%%%%%%%
\section{Conclusions \label{sec:conclusions}}

In this work we have explored the rotational dynamics of anisotropic fluid tracers particles in the Rayleigh-B\'enard flow in two dimensions.  We showed that  elongated particles  align preferentially with the direction of the fluid flow, i.e., horizontally close to the isothermal walls and dominantly vertically in the bulk. This behaviour is due to the large scale circulation flow structure, which induces strong shear at wall boundaries and in up/down-welling regions. In  shear dominated regions the particles performs Jeffery orbits and therefore their rotation rate slows down for orientations parallel to the flow (and orthogonal to the shear direction).  
The near-wall horizontal alignment of rods persists at increasing the Rayleigh number, while the vertical orientation in the bulk is progressively weakened by the corresponding increase of turbulence intensity.
Furthermore, we showed that very elongated particles are nearly orthogonal to the orientation of the temperature gradient, an alignment independent of the system dimensionality and which becomes exact in the limit of infinite Prandtl numbers.  
Tumbling rates are extremely vigorous adjacent to the walls in particular for nearly isotropic particles.  At $Ra=10^9$ the root-mean-square tumbling rate for spheres is $\mathcal{O}(10)$ times stronger than for rods. In the turbulent bulk  the situation reverses and rods tumble slightly faster than isotropic particles, in agreement with earlier observations in two-dimensional turbulence.

Additionally, the tumbling dynamics at the center of the system allows to asses  the level of statistical isotropy of the flow system. 
 It appears that such an isotropy is not yet fully recovered at the highest Rayleigh number simulated in this study ($Ra = 8 \times 10^9$). This suggest the possibility to use rods as a proxy to estimate isotropy in two-dimensional flows.  We have provided a relation that links the tumbling rate to the aspect ratio in case of a statistically isotropic flow. We plan, in a forthcoming study, to extend our investigation to the case of anisotropic particles in a realistic three-dimensional convective system.\\
 
\textit{Acknowledgments} 
The M\'eso-centre de Calcul Scientifique Intensif de l'Universit\'e de Lille (\url{hpc.univ-lille.fr}) is acknowledged for providing computing resources.

\section{APPENDIX} \label{sec:appendix}

\subsection{Equation for the dynamics of the particle orientation angle $\theta$ (or Jeffery equation in 2D)}\label{sec:jeffery-2d}
The dynamics of the orientation unit vector $\textbf{p}$ of a ellipsoidal inertialess axi-symmetric particle in a spatially linear flow is described by the following equation:
\begin{eqnarray}
\dot{\textbf{p}} = \Omega \textbf{p} + \Lambda \left( \mathcal{S}\textbf{p} - (\textbf{p}^{T}\mathcal{S}\textbf{p})\ \textbf{p} \right)
\end{eqnarray}
with 
\begin{eqnarray}
 \Omega &\equiv& \frac{1}{2}\left( \bm{\partial	} \bm{u} - (\bm{\partial} \bm{u})^{T}\right)  , \quad  
 \mathcal{S}   \equiv \frac{1}{2}\left( \bm{\partial} \bm{u} + (\bm{\partial} \bm{u})^{T}\right),\nonumber\\
 \bm{\partial} \bm{u} &=& 
\begin{pmatrix} 
\frac{\partial u_x}{\partial x} & \frac{\partial u_x}{\partial y} & \frac{\partial u_x}{\partial z}\\
\frac{\partial u_y}{\partial x} & \frac{\partial u_y}{\partial y} & \frac{\partial u_y}{\partial z}\\
\frac{\partial u_z}{\partial x} & \frac{\partial u_z}{\partial y} & \frac{\partial u_z}{\partial z}
\end{pmatrix} , \quad
\Lambda = \frac{\alpha^2 - 1}{\alpha^2+1},
\end{eqnarray}
where $\alpha = l/d$ is the length over diameter aspect ratio and $ \bm{\partial} \bm{u}(\textbf{r}(t),t)$ is the fluid velocity gradient tensor at the particle position $\textbf{r}(t)$.
In two-dimension the above equation can be simplified by using the following relations:
\begin{widetext}
\begin{eqnarray}
\textbf{p} &=& 
\begin{pmatrix} 
p_x\\ 
p_y 
\end{pmatrix}  = 
\begin{pmatrix}
\cos{\theta}\\ 
\sin{\theta} 
\end{pmatrix} , \quad \\
\Omega &=& 
\begin{pmatrix} 
0 & \Omega_{xy} \\ 
-\Omega_{xy} & 0 
\end{pmatrix} = 
\begin{pmatrix} 
0 &  \frac{1}{2}\left(  \frac{\partial u_x}{\partial y} -  \frac{\partial u_y}{\partial x}\right) \\ 
  \frac{1}{2}\left(  \frac{\partial u_y}{\partial x} -  \frac{\partial u_x}{\partial y}\right)& 0 
\end{pmatrix} = 
\begin{pmatrix} 
0 & -\omega/2 \\ 
\omega/2 & 0 
\end{pmatrix} , \quad \\
\mathcal{S} &=&
 \begin{pmatrix} 
 S_{xx}& S_{xy} \\ 
 S_{xy} & S_{yy}
\end{pmatrix} =
\begin{pmatrix} 
\frac{\partial u_x}{\partial x}  &  \frac{1}{2}\left(  \frac{\partial u_x}{\partial y} + \frac{\partial u_y}{\partial x}\right) \\ 
  \frac{1}{2}\left(  \frac{\partial u_y}{\partial x}  + \frac{\partial u_x}{\partial y}\right)& \frac{\partial u_y}{\partial y} 
\end{pmatrix} = 
 \begin{pmatrix} 
 S_{xx}& S_{xy} \\ 
 S_{xy} & - S_{xx}
\end{pmatrix}
\end{eqnarray}
\end{widetext}
where $\omega$ is the vorticity pseudo-scalar ( defined as $\omega \hat{z} = \bm{\partial} \times \textbf{u}$, and the relation $S_{yy} = -S_{xx}$ is a direct consequence of the flow incompressibility, $\bm{\partial} \cdot \textbf{u} = 0$. The equations for $p_x$ and $p_y$ are redundant, we just develop the one for the $x$ component:
%%%%%%%%%%
\begin{eqnarray}
%\dot{p_x} &=& \Omega_{xy}\ p_y + \Lambda \left[ S_{xx} p_x + S_{xy} p_y - (p_x S_{xx} p_x + p_x S_{xy}p_y + p_y S_{xy} p_x + p_y S_{yy} p_y)p_x\right]\\
%\dot{p_x} &=& \Omega_{xy}\ p_y + \Lambda \left[ S_{xx} p_x + S_{xy} p_y -  S_{xx} p_x^3 -  S_{xy} p_x^2 p_y -  S_{xy} p_x^2 p_y +  S_{xx} p_x p_y^2 \right]\\
\dot{p_x} &=& \Omega_{xy}\ p_y\nonumber\\
 &+& \Lambda \left[ S_{xx} (p_x -  p_x^3  + p_x p_y^2  ) +  S_{xy} ( p_y - 2 p_x^2 p_y )\right],%\\
\end{eqnarray}
by introducing the angle $\theta$, it becomes:
\begin{eqnarray}
%- \sin(\theta) \dot{\theta} &=&  \Omega_{xy} \sin(\theta) + \Lambda \left[   S_{xx} (\cos{\theta} - \cos{\theta}^3 + \cos{\theta}\sin{\theta}^2 )  + S_{xy} (\sin{\theta} - 2 \cos{\theta}^2\sin{\theta})\right]\\
% \dot{\theta} &=&  - \Omega_{xy}  - \Lambda \left[  S_{xx} (\cot{\theta} - \cot{\theta}\cos{\theta}^2 + \cos{\theta}\sin{\theta} )  + S_{xy} (1 - 2 \cos{\theta}^2)\right]\\
  %\dot{\theta} &=&  - \Omega_{xy}  - \Lambda \left[   S_{xx} (2\cos{\theta}\sin{\theta} )  + S_{xy} (1 - 2 \cos{\theta}^2)\right)\\
   \dot{\theta} &=&  - \Omega_{xy}  -\Lambda \left[   S_{xx} \sin(2\theta)  - S_{xy} \cos(2\theta) \right]%\\
\end{eqnarray}
or
\begin{eqnarray}\label{eq:jef-app}
     \dot{\theta} &=&  \frac{1}{2} \omega  -\Lambda \left[  S_{xx} \sin(2\theta)  - S_{xy} \cos(2\theta) \right].
\end{eqnarray}
The latter equation is consistent with  equation (1) in Parsa \textit{et al.} (2011) \cite{ParsaPF2011}.% (while there is a sign typo in front of $\Lambda$ in \cite{GuptaPRE2014}). 

%%%%%%%%%%%%%%%%%%%%%%%%%%%%%%%%%%%%%%%%%%%%%%%%%%%%%%%%%%%%%%%%%%%%%%%%%%%%%%%%%%%
\subsection{Equation of particle orientation with respect to the rate-of-strain eigenvalues}\label{sec:eq-e1}
The symmetric tensor $\mathcal{S}$ has two orthogonal eigenvectors $\textbf{e}_1$ and $\textbf{e}_2$ and real eigenvalues $\lambda_1, \lambda_2$ which are opposite in sign due to the incompressibility of the flow. 
This means that:
\begin{equation}
\mathcal{S} =  \begin{pmatrix}
S_{xx} & S_{xy}\\ 
S_{xy} & - S_{xx}
\end{pmatrix}
=
(\textbf{e}_1, \textbf{e}_2) 
\begin{pmatrix}
\lambda_1 & 0\\ 
0 & -\lambda_1
\end{pmatrix}
\begin{pmatrix}
\textbf{e}_1\\
\textbf{e}_2
\end{pmatrix}
\end{equation}
where $\lambda_1 = \sqrt{ S_{xx}^2 + S_{xy}^2}$.
%$$\mathcal{S}\textbf{p} = \lambda_1 \left[ (\textbf{e}_1\cdot \textbf{p}) \textbf{e}_1  - (\textbf{e}_2\cdot \textbf{p}) \textbf{e}_2\right]$$
%
%
%$$\textbf{p}^T \mathcal{S}\textbf{p} = \lambda_1 \left[ (\textbf{e}_1\cdot \textbf{p})^2 - (\textbf{e}_2\cdot \textbf{p})^2\right]$$
%
%\begin{eqnarray}
%\dot{\textbf{p}} = \Omega \textbf{p} + \Lambda   \lambda_1 \left(  (\textbf{e}_1\cdot \textbf{p}) \textbf{e}_1  - (\textbf{e}_2\cdot \textbf{p}) \textbf{e}_2  -  \left[ (\textbf{e}_1\cdot \textbf{p})^2 - (\textbf{e}_2\cdot \textbf{p})^2\right]\textbf{p} \right)
%\end{eqnarray}
By introducing 
$
\textbf{e}_1  = 
\begin{pmatrix}
\cos{\theta_1}\\ 
\sin{\theta_1} 
\end{pmatrix}
$ and $
\textbf{e}_2  = 
\begin{pmatrix}
\cos{(\theta_1+\pi/2)}\\ 
\sin{(\theta_1+\pi/2)} 
\end{pmatrix} =
\begin{pmatrix}
- \sin{\theta_1}\\ 
\cos{\theta_1} 
\end{pmatrix}
$
one gets:
%\begin{equation}
%\begin{cases}
%S_{xx} = \sqrt{S_{xx}^2 + S_{xy}^2}\ (2 \cos^2\theta_1 - 1) = \sqrt{S_{xx}^2 + S_{xy}^2}\ \cos(2\theta_1)\\
%
%S_{xy} = \sqrt{S_{xx}^2 + S_{xy}^2}\ 2 \sin\theta_1\cos\theta_1 = \sqrt{S_{xx}^2 + S_{xy}^2}\ \sin(2\theta_1)
%\end{cases}
%\end{equation}
\begin{eqnarray}
S_{xx}  &=& \sqrt{S_{xx}^2 + S_{xy}^2}\ \cos(2\theta_1)\ ,\nonumber\\
S_{xy}  &=& \sqrt{S_{xx}^2 + S_{xy}^2}\ \sin(2\theta_1)
\end{eqnarray}
which can be plugged in into eq. (\ref{eq:jef-app}) to obtain 
\begin{equation}
\dot{\theta} =  \frac{1}{2} \omega  -\frac{\alpha^2 - 1}{\alpha^2+1}\sqrt{ S_{xx}^2 + S_{xy}^2}\ \sin(2(\theta-\theta_1)).
\end{equation} 

%%%%%%%%%%%%%%%%%%%%%%%%%%%%%%%%%%%%%%%%%%%%%%%%%%%%%%%%%%%%%%%%%%%%%%%%
\subsection{Lagrangian equation for the temperature gradient orientation}\label{sec:eq-gradt}
Taking the gradient of the advection diffusion equation for temperature (\ref{eq:T}):
\begin{equation}
\dot{\bm{\partial}T}  =   - (\bm{\partial}\textbf{u})^T \bm{\partial}T  + \kappa\ \partial^2 \bm{\partial}T, 
\end{equation}
where the superscript dot symbol $(\cdot)$ denotes as for the particles the derivative in the Lagrangian reference frame.
The equation for the unit norm vector $\hat{\bm{\partial}T} = \bm{\partial}T / || \bm{\partial}T ||$ is obtained derivation and by taking into account the normalization constraint $(\hat{\bm{\partial}T})^T \hat{\bm{\partial}T} = 1$:

\begin{equation}
\dot{\hat{\bm{\partial}T}}  =   - (\bm{\partial}\textbf{u})^T \hat{\bm{\partial}T}   + (\hat{\bm{\partial}T})^T \mathcal{S}\hat{\bm{\partial}T}  \hat{\bm{\partial}T} + \mathcal{O}(\kappa)
\end{equation}
or:
\begin{equation}
\dot{\hat{\bm{\partial}T}}  =   \Omega \hat{\bm{\partial}T}  - \mathcal{S} \hat{\bm{\partial}T}   + (\hat{\bm{\partial}T})^T \mathcal{S}\hat{\bm{\partial}T}  \hat{\bm{\partial}T} + \mathcal{O}(\kappa),
\end{equation}
where $\mathcal{O}(\kappa)$ denotes the dissipative terms of linear order in $\kappa$. Apart form the dissipative terms, one can immediately remark the strict similarity with the Jeffery equation (\ref{eq:Jeffery3d}) for $\textbf{p}$ in the limit $\alpha\to 0$, which represents the limit of a thin oblate particle (i.e. a disk) in 3D.

The scalar product between the particle orientation (\ref{eq:Jeffery3d}) and $\hat{\bm{\partial}T}$ becomes:
\begin{eqnarray}
\frac{d}{dt}(\textbf{p}^T \hat{\bm{\partial}T} ) &=& (\Lambda - 1) \textbf{p}^T\mathcal{S} \hat{\bm{\partial}T} \nonumber\\
&-& \left[ \Lambda (\textbf{p}^{T}\mathcal{S}\textbf{p})  - (\hat{\bm{\partial}T})^T \mathcal{S}\hat{\bm{\partial}T}  \right]  \textbf{p}^T  \hat{\bm{\partial}T}\nonumber\\
&+& \mathcal{O}(\kappa) 
\end{eqnarray}
which neglecting the terms associated to the dissipation and in the limit $\Lambda \to 1$ ($\alpha \to \infty$) has a fixed point solution for $\textbf{p}^T \hat{\bm{\partial}T} = 0$.
Note also that in dimensionless dissipative units the diffusive terms become proportional to $Pr^{-1}$ meaning that the alignment does not decrease by changing the turbulence intensity ($Ra$ number in this specific case) but it depends on the ratio between diffusive and viscous processes. Therefore exact orthoghonality between $\hat{\bm{\partial}T}$ and $\textbf{p}$ can be reached only in the $Pr \to \infty$ limit.

\subsection{Predictions for the mean quadratic tumbling rate in two-dimensions}\label{sec:tumbl-relations}

The tumbling rate in two dimensions is defined by $\langle \dot{\textbf{p}} \cdot \dot{\textbf{p}} \rangle = \langle \dot{\theta}^2\rangle$.
We begin from:
\begin{equation}
\dot{\theta} =  \tfrac{1}{2} \omega  -\tfrac{\alpha^2-1}{\alpha^2+1} \left[  S_{xx} \sin(2\theta)  - S_{xy} \cos(2\theta) \right]
\end{equation}   

\paragraph{Stationary plane shear flow}
%%%%%%%%%%%%%%
%\begin{wrapfigure}{r}{0.49\textwidth}
\begin{figure}
  \begin{center}
    \includegraphics[width=0.3\textwidth]{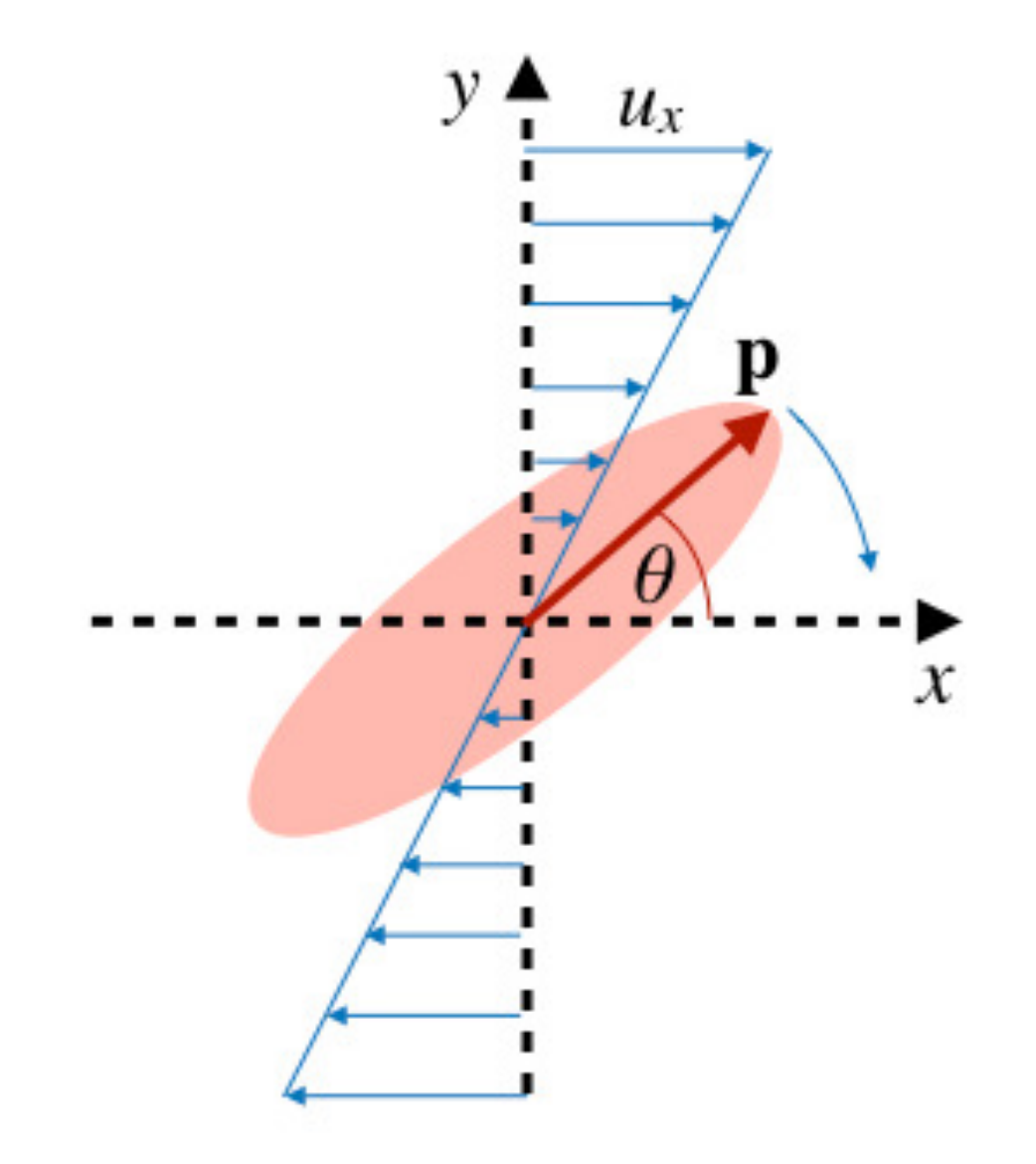}
  \end{center}
  \caption{Anisotropic particle in a plane linear shear flow, \textit{i.e.} $u_x(y)=\dot{\gamma} y$ with $\dot{\gamma}=const.>0$.}\label{fig:cartoon}
%\end{wrapfigure}
\end{figure}
%%%%%%%%%%%%% 
In a plane shear flow $\dot{\gamma} = \partial_y u_x$ the evolution equation for the orientation becomes 
\begin{equation}
\dot{\theta} =  \frac{1}{2} \dot{\gamma}  \left(-1 + \tfrac{\alpha^2-1}{\alpha^2+1}   \cos(2\theta) \right) 
\end{equation}  
with the initial condition $\theta(t_0)= \theta_0$, it  can be solved as
\begin{equation}\label{eq:jsolution}
\tan(\theta)  = \frac{1}{\alpha} \tan\left( \frac{-\dot{\gamma}(t-t_0)}{\alpha+1/\alpha} + \textrm{atan}(\alpha \tan(\theta_0)) \right),
\end{equation}
note that with our shear direction choice the angle decreases with time  (see Fig. \ref{fig:cartoon}). 
The above solution is periodic with a period $T$ required for a rotation of $\pi$: 
\begin{equation}
T = \frac{\pi}{\dot{\gamma}}\left(\alpha+\frac{1}{\alpha}\right).
\end{equation}
By deriving with respect to time (\ref{eq:jsolution}), and setting for simplicity $t_0=0$, $\theta_0=0$, one obtains:
\begin{equation}
\dot{\theta} = \frac{-\gamma \alpha^2}{\alpha^2+1} \frac{\tan^2\left( \frac{-\dot{\gamma}t}{\alpha+1/\alpha} \right) + 1}{\tan^2\left( \frac{-\dot{\gamma}t}{\alpha+1/\alpha} \right) + \alpha^2} = -\frac{\pi\alpha}{T} \frac{\tan^2\left(-\pi \frac{t}{T} \right) + 1}{\tan^2\left( -\pi \frac{t}{T} \right) + \alpha^2}
\end{equation}
which can be squared and averaged over its period T:
\begin{equation}
 \langle\dot{\theta}^2 \rangle = \dot{\gamma}^2 \frac{\alpha}{2(\alpha^2+1)} =  \frac{\epsilon}{\nu}\ \frac{\alpha}{2(\alpha^2+1)}
\end{equation}

\paragraph{Uncorrelated orientation}
We assume that the fluid velocity gradient components are statistically independent and that they are uncorrelated with the particle orientation angle. By squaring and averaging over time and ensembles eq. (\ref{eq:theta}) one obtains:
\begin{eqnarray}
\langle \dot{\theta}^2  \rangle &=&  \frac{1}{4}\langle \omega^2 \rangle + \frac{1}{2}  \left(\frac{\alpha^2-1}{\alpha^2+1} \right)^2\left[  \langle S_{xx}^2 \rangle   +  \langle S_{xy}^2 \rangle \right]
\end{eqnarray} 
\paragraph{Uncorrelated orientation and statistical isotropy and homogeneity}
By taking into account the isotropic relations derived in Sec. \ref{sec:iso}, eq. (\ref{eq:isograd}), we find:  %\textcolor{blue}{we find:
\begin{equation}
\frac{\langle \dot{\theta}^2  \rangle}{\epsilon/\nu} =  \frac{1}{4} + \frac{1}{8} \left(\frac{\alpha^2-1}{\alpha^2+1} \right)^2
\end{equation}

\subsection{Statistical isotropy and homogeneity in two-dimensions}\label{sec:iso}
The present derivation follows the one provided in Ref.  \cite{Pumir2017} for 3D flows. The general form for a fourth order isotropic tensor is:
\begin{equation}
\langle \partial_i u_j \partial_k u_l \rangle = A\ \delta_{ij} \delta_{kl} + B\ \delta_{ik} \delta_{jl} + C\ \delta_{il} \delta_{jk}
\end{equation}
where the indexes $i,j,k,l$ can all independently take the labels $x,y$. The summation over repeated indices is assumed in the following.
The flow incompressibility, $\partial_i u_i=0$, implies that 
\begin{equation}
\langle \partial_i u_i \partial_k u_l \rangle = 0,
\end{equation}
the homogeneity, i.e. statistical translational invariace, of the system instead implies that 
\begin{equation}
\langle \partial_i u_j \partial_j u_i \rangle = 0, 
\end{equation}
and finally the definition of energy dissipation rate is: 
\begin{equation}
\nu \langle \partial_i u_j \partial_i u_j \rangle = \langle \epsilon \rangle.
\end{equation}
The three above equations leads to the system:
\begin{equation}
\begin{cases}
2 A + B + C = 0\\
2 A + 2 B + 4 C = 0\\
2 A + 4 B + 2 C = \langle \epsilon \rangle / \nu
\end{cases}
\end{equation}
which gives as a solution $A=C=-\frac{\langle \epsilon \rangle}{8\nu}$ and $B=\frac{3\langle \epsilon \rangle}{8\nu}$, this leads to:
\begin{equation}
\langle \partial_i u_j \partial_k u_l \rangle = \frac{\langle \epsilon \rangle}{8\nu}\left( 3 \delta_{ik} \delta_{jl} - \delta_{ij} \delta_{kl} -\delta_{il} \delta_{jk}\right)
\end{equation}
and therefore:
\begin{eqnarray}
\langle (\partial_x u_x)^2 \rangle &=& \langle (\partial_y u_y)^2 \rangle = \frac{\langle \epsilon \rangle}{8\nu}\\ 
\langle (\partial_x u_y)^2 \rangle &=& \langle (\partial_y u_x)^2 \rangle = \frac{3\langle \epsilon \rangle}{8\nu}\\
\langle \partial_x u_y \partial_y u_x \rangle &=& -\frac{\langle \epsilon \rangle}{8\nu}
\end{eqnarray}
or also
\begin{equation}\label{eq:isograd}
\langle S_{xx}^2 \rangle = \langle S_{yy}^2 \rangle = \langle S_{xy}^2 \rangle = \frac{\langle \epsilon \rangle}{8\nu}, \quad \langle \omega^2 \rangle = \frac{\langle \epsilon \rangle}{\nu}
\end{equation}

\nocite{apsrev41Control} 
\bibliographystyle{apsrev4-1}
%\bibliography{biblio}{}

\begin{thebibliography}{28}%
\makeatletter
\providecommand \@ifxundefined [1]{%
 \@ifx{#1\undefined}
}%
\providecommand \@ifnum [1]{%
 \ifnum #1\expandafter \@firstoftwo
 \else \expandafter \@secondoftwo
 \fi
}%
\providecommand \@ifx [1]{%
 \ifx #1\expandafter \@firstoftwo
 \else \expandafter \@secondoftwo
 \fi
}%
\providecommand \natexlab [1]{#1}%
\providecommand \enquote  [1]{``#1''}%
\providecommand \bibnamefont  [1]{#1}%
\providecommand \bibfnamefont [1]{#1}%
\providecommand \citenamefont [1]{#1}%
\providecommand \href@noop [0]{\@secondoftwo}%
\providecommand \href [0]{\begingroup \@sanitize@url \@href}%
\providecommand \@href[1]{\@@startlink{#1}\@@href}%
\providecommand \@@href[1]{\endgroup#1\@@endlink}%
\providecommand \@sanitize@url [0]{\catcode `\\12\catcode `\$12\catcode
  `\&12\catcode `\#12\catcode `\^12\catcode `\_12\catcode `\%12\relax}%
\providecommand \@@startlink[1]{}%
\providecommand \@@endlink[0]{}%
\providecommand \url  [0]{\begingroup\@sanitize@url \@url }%
\providecommand \@url [1]{\endgroup\@href {#1}{\urlprefix }}%
\providecommand \urlprefix  [0]{URL }%
\providecommand \Eprint [0]{\href }%
\providecommand \doibase [0]{http://dx.doi.org/}%
\providecommand \selectlanguage [0]{\@gobble}%
\providecommand \bibinfo  [0]{\@secondoftwo}%
\providecommand \bibfield  [0]{\@secondoftwo}%
\providecommand \translation [1]{[#1]}%
\providecommand \BibitemOpen [0]{}%
\providecommand \bibitemStop [0]{}%
\providecommand \bibitemNoStop [0]{.\EOS\space}%
\providecommand \EOS [0]{\spacefactor3000\relax}%
\providecommand \BibitemShut  [1]{\csname bibitem#1\endcsname}%
\let\auto@bib@innerbib\@empty
%</preamble>
\bibitem [{\citenamefont {Voth}\ and\ \citenamefont
  {Soldati}(2017)}]{VothARFM2017}%
  \BibitemOpen
  \bibfield  {author} {\bibinfo {author} {\bibfnamefont {G.~A.}\ \bibnamefont
  {Voth}}\ and\ \bibinfo {author} {\bibfnamefont {A.}~\bibnamefont {Soldati}},\
  }\bibfield  {title} {\enquote {\bibinfo {title} {Anisotropic particles in
  turbulence},}\ }\href {\doibase 10.1146/annurev-fluid-010816-060135}
  {\bibfield  {journal} {\bibinfo  {journal} {Ann. Rev. Fluid Mech.}\ }\textbf
  {\bibinfo {volume} {49}},\ \bibinfo {pages} {249--276} (\bibinfo {year}
  {2017})}\BibitemShut {NoStop}%
\bibitem [{\citenamefont {Parsa}\ \emph {et~al.}(2012)\citenamefont {Parsa},
  \citenamefont {Calzavarini}, \citenamefont {Toschi},\ and\ \citenamefont
  {Voth}}]{ParsaPRL2012}%
  \BibitemOpen
  \bibfield  {author} {\bibinfo {author} {\bibfnamefont {S.}~\bibnamefont
  {Parsa}}, \bibinfo {author} {\bibfnamefont {E.}~\bibnamefont {Calzavarini}},
  \bibinfo {author} {\bibfnamefont {F.}~\bibnamefont {Toschi}}, \ and\ \bibinfo
  {author} {\bibfnamefont {Greg~A.}\ \bibnamefont {Voth}},\ }\bibfield  {title}
  {\enquote {\bibinfo {title} {Rotation rate of rods in turbulent fluid
  flow},}\ }\href {\doibase 10.1103/PhysRevLett.109.134501} {\bibfield
  {journal} {\bibinfo  {journal} {Phys. Rev. Lett.}\ }\textbf {\bibinfo
  {volume} {109}},\ \bibinfo {pages} {134501} (\bibinfo {year}
  {2012})}\BibitemShut {NoStop}%
\bibitem [{\citenamefont {Parsa}\ and\ \citenamefont
  {Voth}(2014)}]{ParsaPRL2014}%
  \BibitemOpen
  \bibfield  {author} {\bibinfo {author} {\bibfnamefont {S.}~\bibnamefont
  {Parsa}}\ and\ \bibinfo {author} {\bibfnamefont {G.~A.}\ \bibnamefont
  {Voth}},\ }\bibfield  {title} {\enquote {\bibinfo {title} {Inertial range
  scaling in rotations of long rods in turbulence},}\ }\href {\doibase
  10.1103/PhysRevLett.112.024501} {\bibfield  {journal} {\bibinfo  {journal}
  {Phys. Rev. Lett.}\ }\textbf {\bibinfo {volume} {112}},\ \bibinfo {pages}
  {024501} (\bibinfo {year} {2014})}\BibitemShut {NoStop}%
\bibitem [{\citenamefont {Marcus}\ \emph {et~al.}(2014)\citenamefont {Marcus},
  \citenamefont {Parsa}, \citenamefont {Kramel}, \citenamefont {Ni},\ and\
  \citenamefont {Voth}}]{Marcus2014}%
  \BibitemOpen
  \bibfield  {author} {\bibinfo {author} {\bibfnamefont {G.~G}\ \bibnamefont
  {Marcus}}, \bibinfo {author} {\bibfnamefont {S.}~\bibnamefont {Parsa}},
  \bibinfo {author} {\bibfnamefont {S.}~\bibnamefont {Kramel}}, \bibinfo
  {author} {\bibfnamefont {R.}~\bibnamefont {Ni}}, \ and\ \bibinfo {author}
  {\bibfnamefont {G.~A.}\ \bibnamefont {Voth}},\ }\bibfield  {title} {\enquote
  {\bibinfo {title} {Measurements of the solid-body rotation of anisotropic
  particles in 3d turbulence},}\ }\href {\doibase
  10.1088/1367-2630/16/10/102001} {\bibfield  {journal} {\bibinfo  {journal}
  {New Journal of Physics}\ }\textbf {\bibinfo {volume} {16}},\ \bibinfo
  {pages} {102001} (\bibinfo {year} {2014})}\BibitemShut {NoStop}%
\bibitem [{\citenamefont {Byron}\ \emph {et~al.}(2015)\citenamefont {Byron},
  \citenamefont {Einarsson}, \citenamefont {Gustavsson}, \citenamefont {Voth},\
  and\ \citenamefont {Mehlig}}]{Byron2015}%
  \BibitemOpen
  \bibfield  {author} {\bibinfo {author} {\bibfnamefont {M.}~\bibnamefont
  {Byron}}, \bibinfo {author} {\bibfnamefont {J.}~\bibnamefont {Einarsson}},
  \bibinfo {author} {\bibfnamefont {K.}~\bibnamefont {Gustavsson}}, \bibinfo
  {author} {\bibfnamefont {G.}~\bibnamefont {Voth}}, \ and\ \bibinfo {author}
  {\bibfnamefont {E.}~\bibnamefont {Mehlig}, \bibfnamefont {B.~andVariano}},\
  }\bibfield  {title} {\enquote {\bibinfo {title} {Shape-dependence of particle
  rotation in isotropic turbulence},}\ }\href@noop {} {\bibfield  {journal}
  {\bibinfo  {journal} {Phys. Fluids}\ }\textbf {\bibinfo {volume} {27}},\
  \bibinfo {pages} {035101} (\bibinfo {year} {2015})}\BibitemShut {NoStop}%
\bibitem [{\citenamefont {Ni}\ \emph {et~al.}(2015)\citenamefont {Ni},
  \citenamefont {Kramel}, \citenamefont {Ouellette},\ and\ \citenamefont
  {Voth}}]{ni_kramel_ouellette_voth_2015}%
  \BibitemOpen
  \bibfield  {author} {\bibinfo {author} {\bibfnamefont {R.}~\bibnamefont
  {Ni}}, \bibinfo {author} {\bibfnamefont {S.}~\bibnamefont {Kramel}}, \bibinfo
  {author} {\bibfnamefont {N.~T.}\ \bibnamefont {Ouellette}}, \ and\ \bibinfo
  {author} {\bibfnamefont {G.~A.}\ \bibnamefont {Voth}},\ }\bibfield  {title}
  {\enquote {\bibinfo {title} {Measurements of the coupling between the
  tumbling of rods and the velocity gradient tensor in turbulence},}\ }\href
  {\doibase 10.1017/jfm.2015.16} {\bibfield  {journal} {\bibinfo  {journal}
  {Journal of Fluid Mechanics}\ }\textbf {\bibinfo {volume} {766}},\ \bibinfo
  {pages} {202–225} (\bibinfo {year} {2015})}\BibitemShut {NoStop}%
\bibitem [{\citenamefont {Bounoua}\ \emph {et~al.}(2018)\citenamefont
  {Bounoua}, \citenamefont {Bouchet},\ and\ \citenamefont
  {Verhille}}]{BounouaPRL2018}%
  \BibitemOpen
  \bibfield  {author} {\bibinfo {author} {\bibfnamefont {S.}~\bibnamefont
  {Bounoua}}, \bibinfo {author} {\bibfnamefont {G.}~\bibnamefont {Bouchet}}, \
  and\ \bibinfo {author} {\bibfnamefont {G.}~\bibnamefont {Verhille}},\
  }\bibfield  {title} {\enquote {\bibinfo {title} {Tumbling of inertial fibers
  in turbulence},}\ }\href {\doibase 10.1103/PhysRevLett.121.124502} {\bibfield
   {journal} {\bibinfo  {journal} {Phys. Rev. Lett.}\ }\textbf {\bibinfo
  {volume} {121}},\ \bibinfo {pages} {124502} (\bibinfo {year}
  {2018})}\BibitemShut {NoStop}%
\bibitem [{\citenamefont {Chevillard}\ and\ \citenamefont
  {Meneveau}(2013)}]{chevillard_meneveau_2013}%
  \BibitemOpen
  \bibfield  {author} {\bibinfo {author} {\bibfnamefont {L.}~\bibnamefont
  {Chevillard}}\ and\ \bibinfo {author} {\bibfnamefont {C.}~\bibnamefont
  {Meneveau}},\ }\bibfield  {title} {\enquote {\bibinfo {title} {Orientation
  dynamics of small, triaxial–ellipsoidal particles in isotropic
  turbulence},}\ }\href {\doibase 10.1017/jfm.2013.580} {\bibfield  {journal}
  {\bibinfo  {journal} {J. Fluid Mechanics}\ }\textbf {\bibinfo {volume}
  {737}},\ \bibinfo {pages} {571–596} (\bibinfo {year} {2013})}\BibitemShut
  {NoStop}%
\bibitem [{\citenamefont {Gustavsson}\ \emph {et~al.}(2014)\citenamefont
  {Gustavsson}, \citenamefont {Einarsson},\ and\ \citenamefont
  {Mehlig}}]{Gustavsson2014}%
  \BibitemOpen
  \bibfield  {author} {\bibinfo {author} {\bibfnamefont {K.}~\bibnamefont
  {Gustavsson}}, \bibinfo {author} {\bibfnamefont {J.}~\bibnamefont
  {Einarsson}}, \ and\ \bibinfo {author} {\bibfnamefont {B.}~\bibnamefont
  {Mehlig}},\ }\bibfield  {title} {\enquote {\bibinfo {title} {Tumbling of
  small axisymmetric particles in random and turbulent flows},}\ }\href@noop {}
  {\bibfield  {journal} {\bibinfo  {journal} {Phys. Rev. Lett.}\ }\textbf
  {\bibinfo {volume} {112}},\ \bibinfo {pages} {014501} (\bibinfo {year}
  {2014})}\BibitemShut {NoStop}%
\bibitem [{\citenamefont {Ni}\ \emph {et~al.}(2014)\citenamefont {Ni},
  \citenamefont {Ouellette},\ and\ \citenamefont
  {Voth}}]{ni_ouellette_voth_2014}%
  \BibitemOpen
  \bibfield  {author} {\bibinfo {author} {\bibfnamefont {R.}~\bibnamefont
  {Ni}}, \bibinfo {author} {\bibfnamefont {N.~T.}\ \bibnamefont {Ouellette}}, \
  and\ \bibinfo {author} {\bibfnamefont {G.~A.}\ \bibnamefont {Voth}},\
  }\bibfield  {title} {\enquote {\bibinfo {title} {Alignment of vorticity and
  rods with lagrangian fluid stretching in turbulence},}\ }\href {\doibase
  10.1017/jfm.2014.32} {\bibfield  {journal} {\bibinfo  {journal} {J. Fluid
  Mechanics}\ }\textbf {\bibinfo {volume} {743}},\ \bibinfo {pages} {R3}
  (\bibinfo {year} {2014})}\BibitemShut {NoStop}%
\bibitem [{\citenamefont {Candelier}\ \emph {et~al.}(2016)\citenamefont
  {Candelier}, \citenamefont {Einarsson},\ and\ \citenamefont
  {Mehlig}}]{CandelierPRL2016}%
  \BibitemOpen
  \bibfield  {author} {\bibinfo {author} {\bibfnamefont {F.}~\bibnamefont
  {Candelier}}, \bibinfo {author} {\bibfnamefont {J.}~\bibnamefont
  {Einarsson}}, \ and\ \bibinfo {author} {\bibfnamefont {B.}~\bibnamefont
  {Mehlig}},\ }\bibfield  {title} {\enquote {\bibinfo {title} {Angular dynamics
  of a small particle in turbulence},}\ }\href {\doibase
  10.1103/PhysRevLett.117.204501} {\bibfield  {journal} {\bibinfo  {journal}
  {Phys. Rev. Lett.}\ }\textbf {\bibinfo {volume} {117}},\ \bibinfo {pages}
  {204501} (\bibinfo {year} {2016})}\BibitemShut {NoStop}%
\bibitem [{\citenamefont {Pujara}\ and\ \citenamefont
  {Variano}(2017)}]{pujara_variano_2017}%
  \BibitemOpen
  \bibfield  {author} {\bibinfo {author} {\bibfnamefont {N.}~\bibnamefont
  {Pujara}}\ and\ \bibinfo {author} {\bibfnamefont {E.~A.}\ \bibnamefont
  {Variano}},\ }\bibfield  {title} {\enquote {\bibinfo {title} {Rotations of
  small, inertialess triaxial ellipsoids in isotropic turbulence},}\ }\href
  {\doibase 10.1017/jfm.2017.256} {\bibfield  {journal} {\bibinfo  {journal}
  {J. Fluid Mechanics}\ }\textbf {\bibinfo {volume} {821}},\ \bibinfo {pages}
  {517–538} (\bibinfo {year} {2017})}\BibitemShut {NoStop}%
\bibitem [{\citenamefont {Gustavsson}\ \emph {et~al.}(2017)\citenamefont
  {Gustavsson}, \citenamefont {Jucha}, \citenamefont {Naso}, \citenamefont
  {L\'ev\^eque}, \citenamefont {Pumir},\ and\ \citenamefont
  {Mehlig}}]{GustavssonPRL2017}%
  \BibitemOpen
  \bibfield  {author} {\bibinfo {author} {\bibfnamefont {K.}~\bibnamefont
  {Gustavsson}}, \bibinfo {author} {\bibfnamefont {J.}~\bibnamefont {Jucha}},
  \bibinfo {author} {\bibfnamefont {A.}~\bibnamefont {Naso}}, \bibinfo {author}
  {\bibfnamefont {E.}~\bibnamefont {L\'ev\^eque}}, \bibinfo {author}
  {\bibfnamefont {A.}~\bibnamefont {Pumir}}, \ and\ \bibinfo {author}
  {\bibfnamefont {B.}~\bibnamefont {Mehlig}},\ }\bibfield  {title} {\enquote
  {\bibinfo {title} {Statistical model for the orientation of nonspherical
  particles settling in turbulence},}\ }\href {\doibase
  10.1103/PhysRevLett.119.254501} {\bibfield  {journal} {\bibinfo  {journal}
  {Phys. Rev. Lett.}\ }\textbf {\bibinfo {volume} {119}},\ \bibinfo {pages}
  {254501} (\bibinfo {year} {2017})}\BibitemShut {NoStop}%
\bibitem [{\citenamefont {Lin}\ \emph {et~al.}(2003)\citenamefont {Lin},
  \citenamefont {Shi},\ and\ \citenamefont {Yu}}]{Lin2003}%
  \BibitemOpen
  \bibfield  {author} {\bibinfo {author} {\bibfnamefont {Jianzhong}\
  \bibnamefont {Lin}}, \bibinfo {author} {\bibfnamefont {Xing}\ \bibnamefont
  {Shi}}, \ and\ \bibinfo {author} {\bibfnamefont {Zhaosheng}\ \bibnamefont
  {Yu}},\ }\bibfield  {title} {\enquote {\bibinfo {title} {The motion of fibers
  in an evolving mixing layer},}\ }\href {\doibase
  https://doi.org/10.1016/S0301-9322(03)00086-7} {\bibfield  {journal}
  {\bibinfo  {journal} {Int. J. Multiphase Flow}\ }\textbf {\bibinfo {volume}
  {29}},\ \bibinfo {pages} {1355 -- 1372} (\bibinfo {year} {2003})}\BibitemShut
  {NoStop}%
\bibitem [{\citenamefont {Lin}\ \emph {et~al.}(2012)\citenamefont {Lin},
  \citenamefont {Liang},\ and\ \citenamefont {Zhang}}]{Lin2012}%
  \BibitemOpen
  \bibfield  {author} {\bibinfo {author} {\bibfnamefont {J.Z.}\ \bibnamefont
  {Lin}}, \bibinfo {author} {\bibfnamefont {X.Y.}\ \bibnamefont {Liang}}, \
  and\ \bibinfo {author} {\bibfnamefont {S.L.}\ \bibnamefont {Zhang}},\
  }\bibfield  {title} {\enquote {\bibinfo {title} {Numerical simulation of
  fiber orientation distribution in round turbulent jet of fiber suspension},}\
  }\href {\doibase https://doi.org/10.1016/j.cherd.2011.09.016} {\bibfield
  {journal} {\bibinfo  {journal} {Chemical Engineering Research and Design}\
  }\textbf {\bibinfo {volume} {90}},\ \bibinfo {pages} {766 -- 775} (\bibinfo
  {year} {2012})},\ \bibinfo {note} {special Issue on the 3rd European Process
  intensification Conference}\BibitemShut {NoStop}%
\bibitem [{\citenamefont {Zhang}\ \emph {et~al.}(2005)\citenamefont {Zhang},
  \citenamefont {Lin},\ and\ \citenamefont {Chan}}]{Lin2005}%
  \BibitemOpen
  \bibfield  {author} {\bibinfo {author} {\bibfnamefont {Ling-Xin}\
  \bibnamefont {Zhang}}, \bibinfo {author} {\bibfnamefont {Jian-Zhong}\
  \bibnamefont {Lin}}, \ and\ \bibinfo {author} {\bibfnamefont {T.~L.}\
  \bibnamefont {Chan}},\ }\bibfield  {title} {\enquote {\bibinfo {title}
  {Orientation distribution of cylindrical particles suspended in a turbulent
  pipe flow},}\ }\href {\doibase 10.1063/1.2046713} {\bibfield  {journal}
  {\bibinfo  {journal} {Physics of Fluids}\ }\textbf {\bibinfo {volume} {17}},\
  \bibinfo {pages} {093105} (\bibinfo {year} {2005})}\BibitemShut {NoStop}%
\bibitem [{\citenamefont {Marchioli}\ \emph {et~al.}(2010)\citenamefont
  {Marchioli}, \citenamefont {Fantoni},\ and\ \citenamefont
  {Soldati}}]{MarchioliPF2010}%
  \BibitemOpen
  \bibfield  {author} {\bibinfo {author} {\bibfnamefont {C.}~\bibnamefont
  {Marchioli}}, \bibinfo {author} {\bibfnamefont {M.}~\bibnamefont {Fantoni}},
  \ and\ \bibinfo {author} {\bibfnamefont {A.}~\bibnamefont {Soldati}},\
  }\bibfield  {title} {\enquote {\bibinfo {title} {Orientation, distribution,
  and deposition of elongated, inertial fibers in turbulent channel flow},}\
  }\href@noop {} {\bibfield  {journal} {\bibinfo  {journal} {Phys. Fluids}\
  }\textbf {\bibinfo {volume} {22}},\ \bibinfo {pages} {033301} (\bibinfo
  {year} {2010})}\BibitemShut {NoStop}%
\bibitem [{\citenamefont {Marchioli}\ and\ \citenamefont
  {Soldati}(2013)}]{Marchioli2013}%
  \BibitemOpen
  \bibfield  {author} {\bibinfo {author} {\bibfnamefont {C.}~\bibnamefont
  {Marchioli}}\ and\ \bibinfo {author} {\bibfnamefont {A.}~\bibnamefont
  {Soldati}},\ }\bibfield  {title} {\enquote {\bibinfo {title} {Rotation
  statistics of fibers in wall shear turbulence},}\ }\href {\doibase
  10.1007/s00707-013-0933-z} {\bibfield  {journal} {\bibinfo  {journal} {Acta
  Mechanica}\ }\textbf {\bibinfo {volume} {224}},\ \bibinfo {pages}
  {2311--2329} (\bibinfo {year} {2013})}\BibitemShut {NoStop}%
\bibitem [{\citenamefont {Zhao}\ \emph {et~al.}(2015)\citenamefont {Zhao},
  \citenamefont {Challabotla}, \citenamefont {Andersson},\ and\ \citenamefont
  {Variano}}]{Zhao2015}%
  \BibitemOpen
  \bibfield  {author} {\bibinfo {author} {\bibfnamefont {L.}~\bibnamefont
  {Zhao}}, \bibinfo {author} {\bibfnamefont {N.~R.}\ \bibnamefont
  {Challabotla}}, \bibinfo {author} {\bibfnamefont {H.~I.}\ \bibnamefont
  {Andersson}}, \ and\ \bibinfo {author} {\bibfnamefont {E.~A.}\ \bibnamefont
  {Variano}},\ }\bibfield  {title} {\enquote {\bibinfo {title} {Rotation of
  non-spherical particles in turbulent channel flow},}\ }\href@noop {}
  {\bibfield  {journal} {\bibinfo  {journal} {Phys. Rev. Lett.}\ }\textbf
  {\bibinfo {volume} {115}},\ \bibinfo {pages} {244501} (\bibinfo {year}
  {2015})}\BibitemShut {NoStop}%
\bibitem [{\citenamefont {Challabotla}\ \emph {et~al.}(2015)\citenamefont
  {Challabotla}, \citenamefont {Zhao},\ and\ \citenamefont
  {Andersson}}]{Challabotla2015}%
  \BibitemOpen
  \bibfield  {author} {\bibinfo {author} {\bibfnamefont {N.R.}\ \bibnamefont
  {Challabotla}}, \bibinfo {author} {\bibfnamefont {L.}~\bibnamefont {Zhao}}, \
  and\ \bibinfo {author} {\bibfnamefont {H.I.}\ \bibnamefont {Andersson}},\
  }\bibfield  {title} {\enquote {\bibinfo {title} {Orientation and rotation of
  inertial disk particles in wall turbulence},}\ }\href@noop {} {\bibfield
  {journal} {\bibinfo  {journal} {J. Fluid Mech.}\ }\textbf {\bibinfo {volume}
  {766}},\ \bibinfo {pages} {R2} (\bibinfo {year} {2015})}\BibitemShut
  {NoStop}%
\bibitem [{\citenamefont {Bakhuis}\ \emph {et~al.}(2019)\citenamefont
  {Bakhuis}, \citenamefont {Mathai}, \citenamefont {Verschoof}, \citenamefont
  {Ezeta}, \citenamefont {Lohse}, \citenamefont {Huisman},\ and\ \citenamefont
  {Sun}}]{Bakhuis2019}%
  \BibitemOpen
  \bibfield  {author} {\bibinfo {author} {\bibfnamefont {D.}~\bibnamefont
  {Bakhuis}}, \bibinfo {author} {\bibfnamefont {V.}~\bibnamefont {Mathai}},
  \bibinfo {author} {\bibfnamefont {R.~A.}\ \bibnamefont {Verschoof}}, \bibinfo
  {author} {\bibfnamefont {R.}~\bibnamefont {Ezeta}}, \bibinfo {author}
  {\bibfnamefont {D.}~\bibnamefont {Lohse}}, \bibinfo {author} {\bibfnamefont
  {S.~G.}\ \bibnamefont {Huisman}}, \ and\ \bibinfo {author} {\bibfnamefont
  {C.}~\bibnamefont {Sun}},\ }\bibfield  {title} {\enquote {\bibinfo {title}
  {Statistics of rigid fibers in strongly sheared turbulence},}\ }\href@noop {}
  {\bibfield  {journal} {\bibinfo  {journal} {Phys. Rev. Fluids}\ }\textbf
  {\bibinfo {volume} {4}},\ \bibinfo {pages} {072301(R)} (\bibinfo {year}
  {2019})}\BibitemShut {NoStop}%
\bibitem [{\citenamefont {Parsa}\ \emph {et~al.}(2011)\citenamefont {Parsa},
  \citenamefont {Guasto}, \citenamefont {Kishore}, \citenamefont {Ouellette},
  \citenamefont {Gollub},\ and\ \citenamefont {Voth}}]{ParsaPF2011}%
  \BibitemOpen
  \bibfield  {author} {\bibinfo {author} {\bibfnamefont {S.}~\bibnamefont
  {Parsa}}, \bibinfo {author} {\bibfnamefont {J.~S.}\ \bibnamefont {Guasto}},
  \bibinfo {author} {\bibfnamefont {M.}~\bibnamefont {Kishore}}, \bibinfo
  {author} {\bibfnamefont {N.~T.}\ \bibnamefont {Ouellette}}, \bibinfo {author}
  {\bibfnamefont {J.~P.}\ \bibnamefont {Gollub}}, \ and\ \bibinfo {author}
  {\bibfnamefont {G.~A.}\ \bibnamefont {Voth}},\ }\bibfield  {title} {\enquote
  {\bibinfo {title} {Rotation and alignment of rods in two-dimensional chaotic
  flow},}\ }\href {\doibase 10.1063/1.3570526} {\bibfield  {journal} {\bibinfo
  {journal} {Phys. Fluids}\ }\textbf {\bibinfo {volume} {23}},\ \bibinfo
  {pages} {043302} (\bibinfo {year} {2011})}\BibitemShut {NoStop}%
\bibitem [{\citenamefont {Gupta}\ \emph {et~al.}(2014)\citenamefont {Gupta},
  \citenamefont {Vincenzi},\ and\ \citenamefont {Pandit}}]{GuptaPRE2014}%
  \BibitemOpen
  \bibfield  {author} {\bibinfo {author} {\bibfnamefont {A.}~\bibnamefont
  {Gupta}}, \bibinfo {author} {\bibfnamefont {D.}~\bibnamefont {Vincenzi}}, \
  and\ \bibinfo {author} {\bibfnamefont {R.}~\bibnamefont {Pandit}},\
  }\bibfield  {title} {\enquote {\bibinfo {title} {Elliptical tracers in
  two-dimensional, homogeneous, isotropic fluid turbulence: The statistics of
  alignment, rotation, and nematic order},}\ }\href {\doibase
  10.1103/PhysRevE.89.021001} {\bibfield  {journal} {\bibinfo  {journal} {Phys.
  Rev. E}\ }\textbf {\bibinfo {volume} {89}},\ \bibinfo {pages} {021001}
  (\bibinfo {year} {2014})}\BibitemShut {NoStop}%
\bibitem [{\citenamefont {Jeffery}(1922)}]{Jeffery1922}%
  \BibitemOpen
  \bibfield  {author} {\bibinfo {author} {\bibfnamefont {G.~B.}\ \bibnamefont
  {Jeffery}},\ }\bibfield  {title} {\enquote {\bibinfo {title} {The motion of
  ellipsoidal particles immersed in a viscous fluid},}\ }\href {\doibase
  10.1098/rspa.1922.0078} {\bibfield  {journal} {\bibinfo  {journal}
  {Proceedings of the Royal Society of London. Series A}\ }\textbf {\bibinfo
  {volume} {102}},\ \bibinfo {pages} {161--179} (\bibinfo {year}
  {1922})}\BibitemShut {NoStop}%
\bibitem [{\citenamefont {Calzavarini}(2019)}]{Calzavarini2019}%
  \BibitemOpen
  \bibfield  {author} {\bibinfo {author} {\bibfnamefont {E.}~\bibnamefont
  {Calzavarini}},\ }\bibfield  {title} {\enquote {\bibinfo {title}
  {Eulerian-lagrangian fluid dynamics platform: The ch4-project},}\ }\href
  {\doibase https://doi.org/10.1016/j.simpa.2019.100002} {\bibfield  {journal}
  {\bibinfo  {journal} {Software Impacts}\ }\textbf {\bibinfo {volume} {1}},\
  \bibinfo {pages} {100002} (\bibinfo {year} {2019})}\BibitemShut {NoStop}%
\bibitem [{\citenamefont {Zhao}\ \emph {et~al.}(2019)\citenamefont {Zhao},
  \citenamefont {Gustavsson}, \citenamefont {Ni}, \citenamefont {Kramel},
  \citenamefont {Voth}, \citenamefont {Andersson},\ and\ \citenamefont
  {Mehlig}}]{Zhao2019}%
  \BibitemOpen
  \bibfield  {author} {\bibinfo {author} {\bibfnamefont {L.}~\bibnamefont
  {Zhao}}, \bibinfo {author} {\bibfnamefont {K.}~\bibnamefont {Gustavsson}},
  \bibinfo {author} {\bibfnamefont {R.}~\bibnamefont {Ni}}, \bibinfo {author}
  {\bibfnamefont {S.}~\bibnamefont {Kramel}}, \bibinfo {author} {\bibfnamefont
  {G.~A.}\ \bibnamefont {Voth}}, \bibinfo {author} {\bibfnamefont {H.~I.}\
  \bibnamefont {Andersson}}, \ and\ \bibinfo {author} {\bibfnamefont
  {B.}~\bibnamefont {Mehlig}},\ }\bibfield  {title} {\enquote {\bibinfo {title}
  {Passive directors in turbulence},}\ }\href {\doibase
  10.1103/PhysRevFluids.4.054602} {\bibfield  {journal} {\bibinfo  {journal}
  {Phys. Rev. Fluids}\ }\textbf {\bibinfo {volume} {4}},\ \bibinfo {pages}
  {054602} (\bibinfo {year} {2019})}\BibitemShut {NoStop}%
\bibitem [{\citenamefont {Pumir}\ and\ \citenamefont
  {Wilkinson}(2011)}]{PumirWilkinson2011}%
  \BibitemOpen
  \bibfield  {author} {\bibinfo {author} {\bibfnamefont {A.}~\bibnamefont
  {Pumir}}\ and\ \bibinfo {author} {\bibfnamefont {M.}~\bibnamefont
  {Wilkinson}},\ }\bibfield  {title} {\enquote {\bibinfo {title} {Orientation
  statistics of small particles in turbulence},}\ }\href {\doibase
  10.1088/1367-2630/13/9/093030} {\bibfield  {journal} {\bibinfo  {journal}
  {New Journal of Physics}\ }\textbf {\bibinfo {volume} {13}},\ \bibinfo
  {pages} {093030} (\bibinfo {year} {2011})}\BibitemShut {NoStop}%
\bibitem [{\citenamefont {Pumir}(2017)}]{Pumir2017}%
  \BibitemOpen
  \bibfield  {author} {\bibinfo {author} {\bibfnamefont {A.}~\bibnamefont
  {Pumir}},\ }\bibfield  {title} {\enquote {\bibinfo {title} {Structure of the
  velocity gradient tensor in turbulent shear flows},}\ }\href {\doibase
  10.1103/PhysRevFluids.2.074602} {\bibfield  {journal} {\bibinfo  {journal}
  {Phys. Rev. Fluids}\ }\textbf {\bibinfo {volume} {2}},\ \bibinfo {pages}
  {074602} (\bibinfo {year} {2017})}\BibitemShut {NoStop}%
\end{thebibliography}

%merlin.mbs apsrev4-1.bst 2010-07-25 4.21a (PWD, AO, DPC) hacked
%Control: key (0)
%Control: author (0) dotless jnrlst
%Control: editor formatted (1) identically to author
%Control: production of article title (0) allowed
%Control: page (1) range
%Control: year (0) verbatim
%Control: production of eprint (0) enabled
%

%%%%%%%%%%%%%%%%%%
%%%%%%%%%%%%%%%%%%

%%%%%%%%%%%%% %%%%%%%%%%%%%%%%%%%ADDITIONAL MATERIAL
\clearpage
%\newpage
%\mbox{}
%\clearpage
\onecolumngrid

\begin{center}
{\large \textbf{Anisotropic particles in two-dimensional convective turbulence,\\
          Additional Material}}\\

Enrico Calzavarini\\
\textit{Univ. Lille, Unit\'e de M\'ecanique de Lille, J. Boussinesq, UML EA 7512, F 59000 Lille, France\\ 
enrico.calzavarini@polytech-lille.fr}\\
Linfeng Jiang, Chao Sun\\
\textit{Center for Combustion Energy, Key Laboratory for Thermal Science and Power Engineering of Ministry of Education,
Department of Energy and Power Engineering, Tsinghua University, Beijing, China}\\
\date{\today}

\end{center}

\appendix
%%%%%%%%%% Prefix a "S" to all equations, figures, tables and reset the counter %%%%%%%%%%
\setcounter{equation}{0}
\setcounter{figure}{0}
\setcounter{table}{0}
\setcounter{page}{1}
\makeatletter

\renewcommand\thefigure{AM.\arabic{figure}}    
%\section*{Additional Material}
\subsection*{Probability density function of nematic order parameter}
\begin{figure}[!ht]
\begin{center}
\includegraphics[width=0.8\columnwidth]{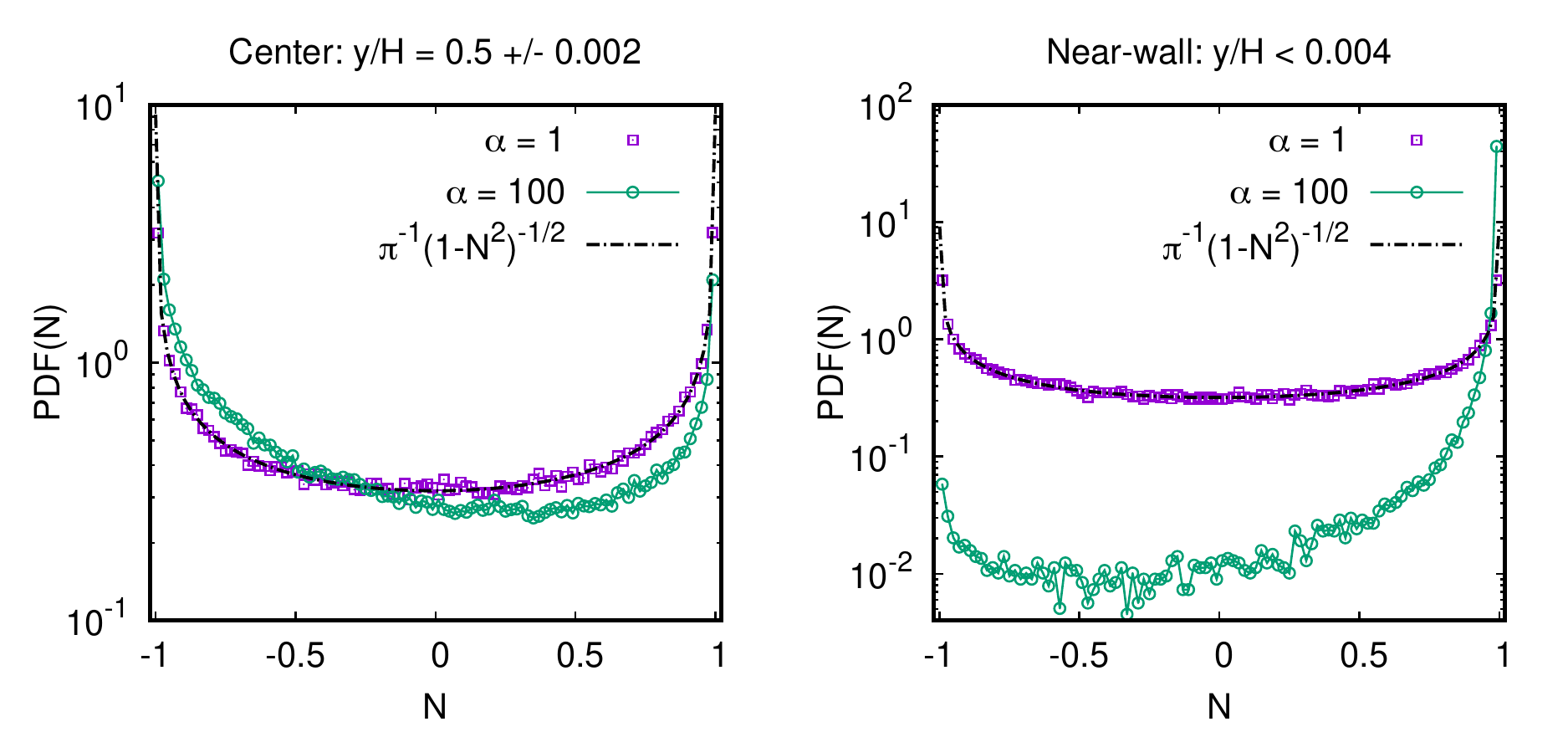}
\caption{Probability density function of the nematic oder parameter $N$ for isotropic ($\alpha = 1$) and highly-anisotropic particles ($\alpha=100$), evaluated close to the center line of the system (left) and close to the wall (right).
 The Rayleigh number is $Ra = 10^9$, and corresponds to the measurements of Fig. 2 of the paper. 
We also report the analytical prediction for the case of homogeneously oriented particles: $PDF(N) = 1/(\pi\sqrt{1-N^2})$.}\label{fig:pdfN}
\end{center}
\end{figure}

In order to support the good convergence of the measurements reported in Fig. 2 of the main manuscript, we report the local (in space) probability density function (PDF) of the nematic order parameter $N$ for two selected cases (Fig. \ref{fig:pdfN}).
%%%%%%%%%% 
%%%%%%%%%%
\subsection*{Results at $Ra=8\times 10^9$, $Pr=1$}
We provide numerical results at the highest Rayleigh number numerically explored in this study, $Ra=8\times 10^9$. These measurements are generally less statistically converged than the lower $Ra$ cases due to the heavier computational costs. However,  they allow to appreciate the increased isotropization of the bulk flow and its consequences on the preferential orientation and tumbling-rate in the bulk, which are in agreement with the phenomenology discussed in the article. 
 
%%%%%%%%%%%%%%%%%
\begin{figure}[!htb]
\begin{center}
\includegraphics[width=0.8\columnwidth]{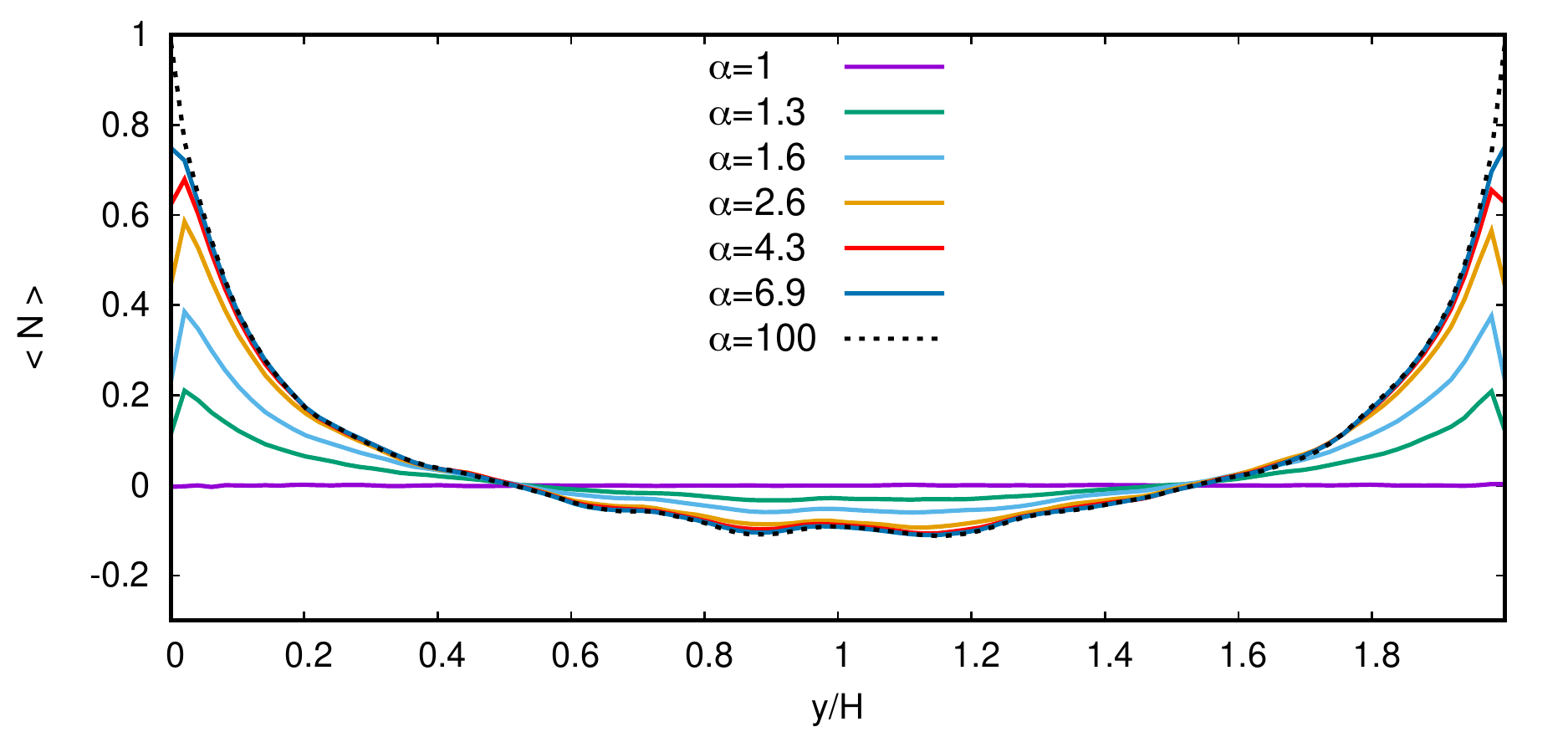}
\caption{Local nematic order parameter as a function the distance from a horizontal wall in the system, for different particles aspect ratios at $Ra=8 \times10^9$, $Pr=1$. We compute the average $\langle N \rangle (y)$, where $\langle \ldots \rangle$ is taken over time and over the particles with given $y \pm \delta y$ coordinates, here with $ \delta y = H/4096$.}\label{fig:nematic-highRa}
\end{center}
\end{figure}
%%%%%%%%%%%%%%%%%%
%%%%%%%%%%%%%%%%%
\begin{figure}[!htb]
\begin{center}
\includegraphics[width=0.7\columnwidth]{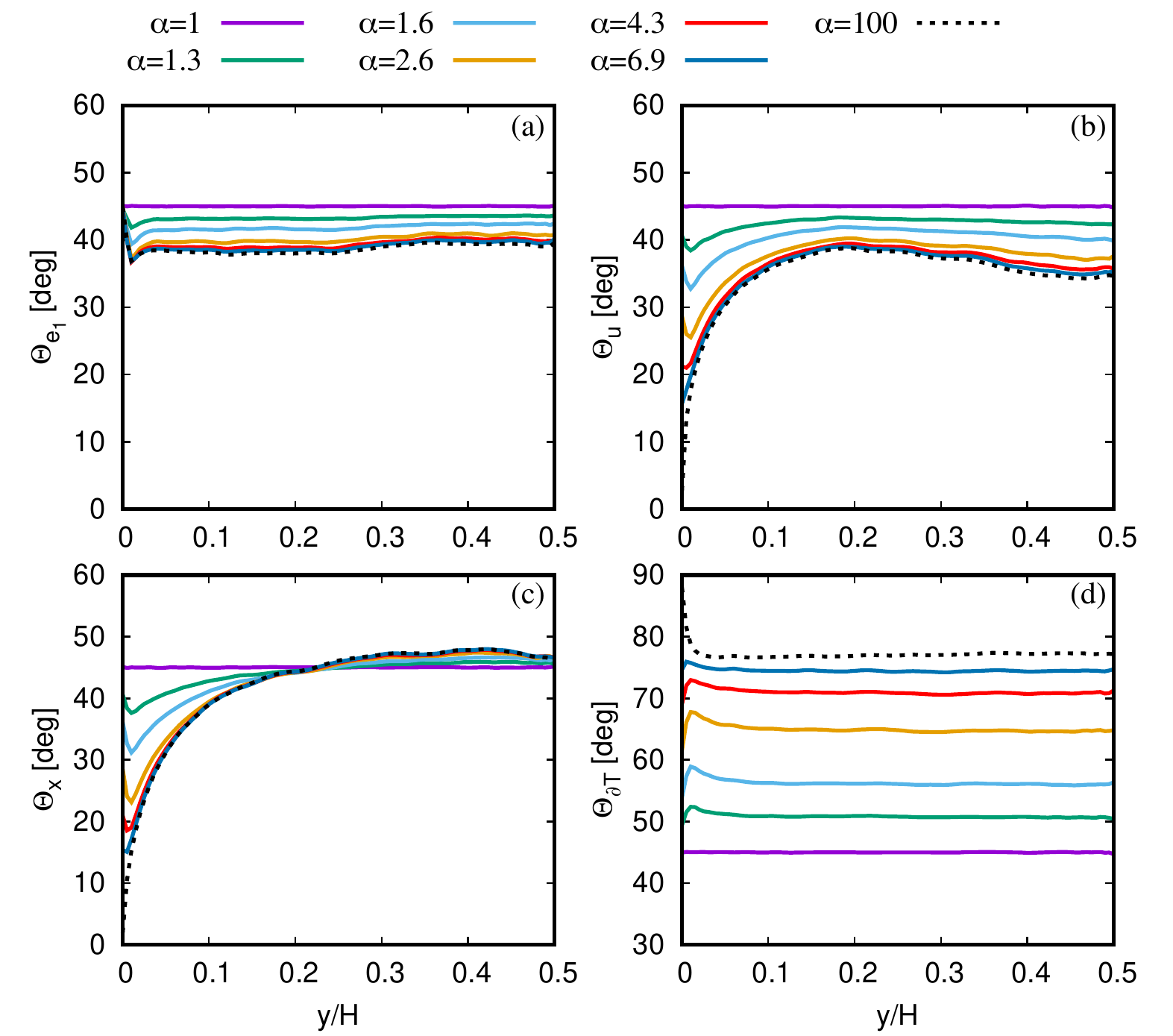}
\caption{Mean orientation angle with respect to the first eigenvector of the rate-of-strain tensor $\textbf{e}_1$ (a); the fluid velocity vector $\textbf{u}$ (b); the $x$ axis (c); the temperature gradient $\bm{\partial}T$ (d), for various particle aspect ratios ranging from spheres $\alpha=1$ to rods $\alpha=100$. $Ra=8 \times 10^9$, $Pr=1$.}\label{fig:orientation-highRa}
\end{center}
\end{figure}
%%%%%%%%%%%%%%
\begin{figure}[!htb]
\begin{center}
\includegraphics[width=0.6\columnwidth]{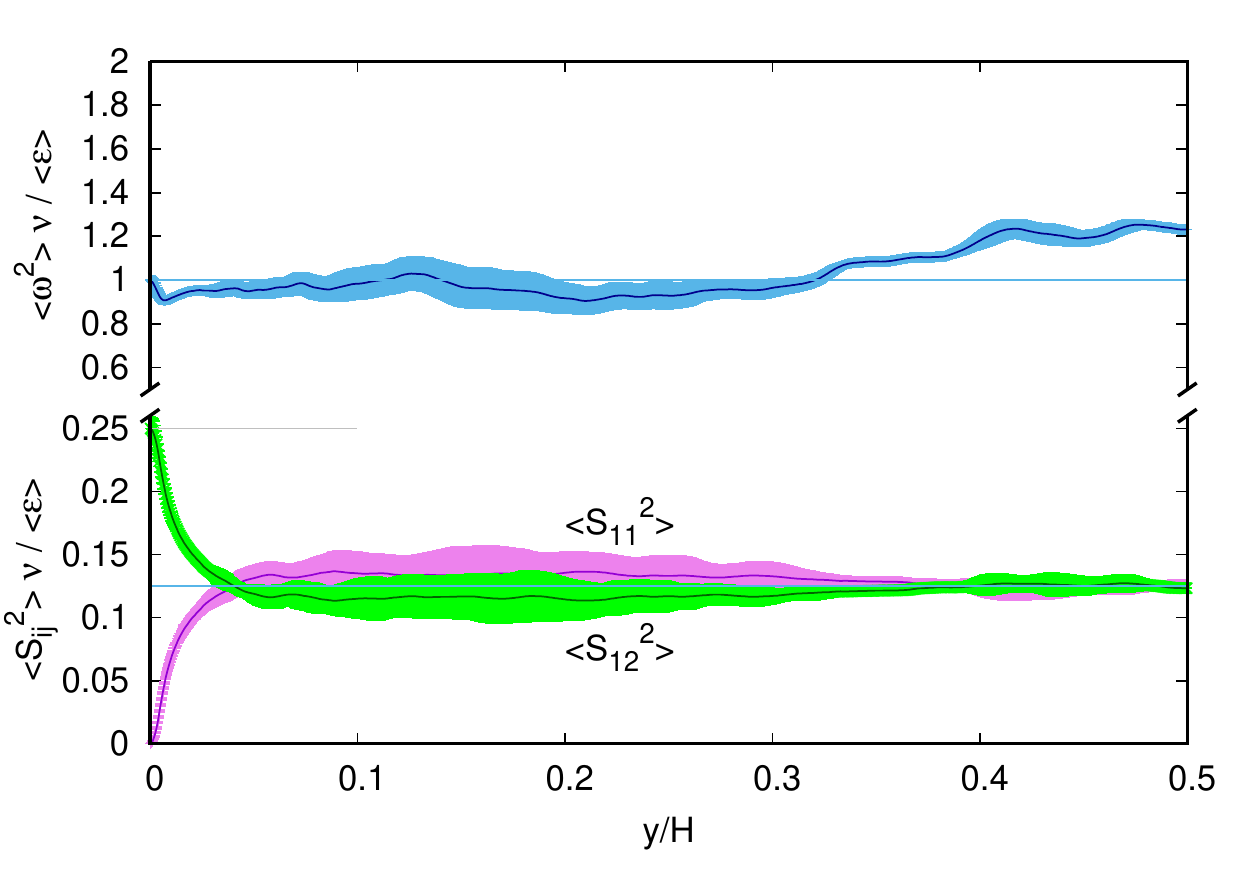}
\caption{Check of local small-scale flow isotropy: same as Fig.6 of the main paper here at $Ra=8\times10^9$, $Pr=1$.  The continuous lines represent $\langle \omega^2 \rangle$,$\langle S_{xx}^2 \rangle$ and $\langle S_{xy}^2 \rangle$ in $\langle \epsilon \rangle / \nu$ units (i.e. local dissipative units) as a function of the distance from the wall $y \in \left[ 0,H \right]$. The colour shadow around the lines indicates the standard deviation error bars.
The dashed lines provides the values expected in the isotropic case, $\langle \omega^2 \rangle \nu / \langle \epsilon \rangle = 1$ and $\langle S_{xx}^2 \rangle \nu / \langle \epsilon \rangle =  \langle S_{xy}^2 \rangle \nu / \langle \epsilon \rangle = 1/8$.   The dotted line reports the value expected for plane shear flow, when the only non-null velocity gradient component is $\partial_y u_x$.}
\label{fig:isotropy-highRa}
\end{center}
\end{figure}
%%%%%%%%%%%%% 
%%%%%%%%%%%%%%%%%
\begin{figure}[!htb]
\begin{center}
\subfigure[]{\includegraphics[width=0.495\columnwidth]{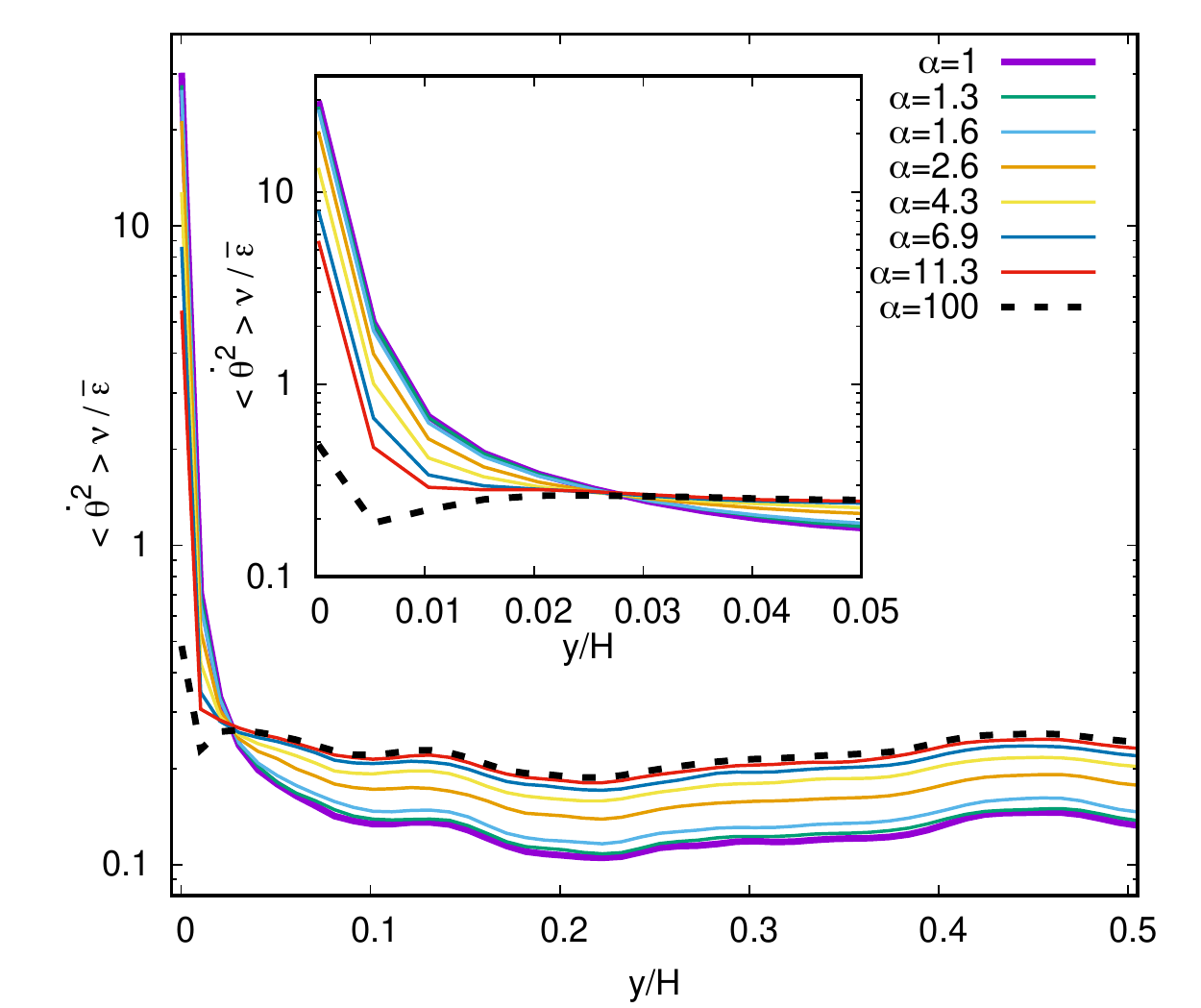}}
\subfigure[]{\includegraphics[width=0.495\columnwidth]{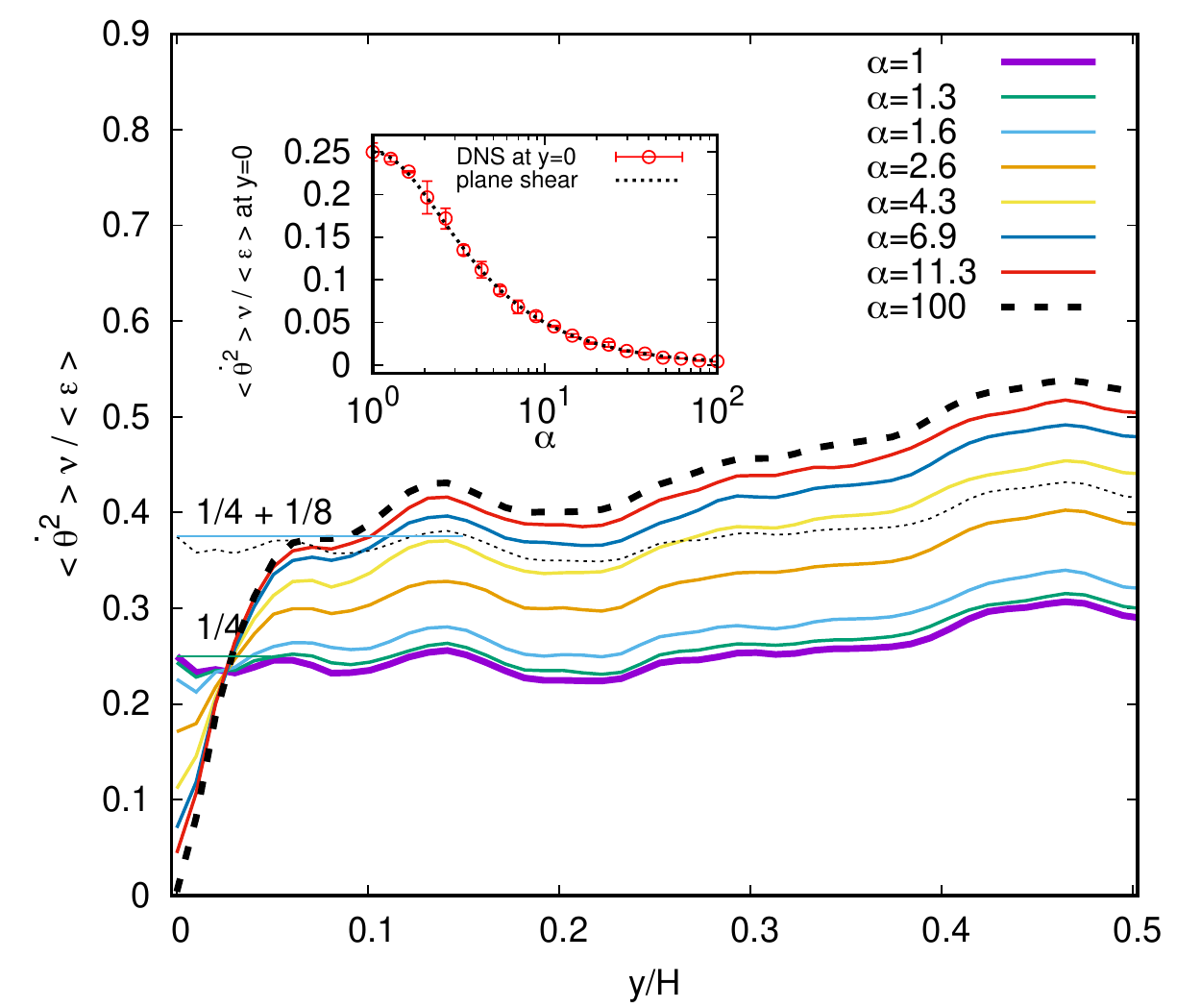}}
\caption{$Ra=8 \times 10^9$ (a) Mean quadratic tumbling rate, $\langle \dot{\theta}^2 \rangle $ as a function of the distance from the wall $y\in [0,H/2]$ for different particle aspect ratios. The tumbling rate is normalized by means of the global energy dissipation rate $\overline{\epsilon}$. The inset reports a zoomed-in vision of the the wall region.
(b) Same as before but with a normalization based on the local dissipative energy dissipation rate $\langle \epsilon \rangle$. The dotted line reports the no-correlation prediction eq. (11)  for $\alpha=100$, the continuous horizontal lines gives the values of the isotropic flow  prediction eq. (12) for $\alpha=1$ (minimum value) and $\alpha=100$(maximum value). The inset reports the values (datapoints) of the normalized quadratic tumbling rate at the wall ($y = 0$) and a comparison with the prediction eq. (10), which describe the tumbling in a plane shear flow.}
\label{fig:tumbling-rate-highRa}
\end{center}
\end{figure}
%%%%%%%%%%%%%%%%%%
%%%%%%%%%%%%%%%%%%

\end{document}